# Tractable Triangles and Cross-Free Convexity
# in Discrete Optimisation

**Martin C. Cooper**                                    COOPER@IRIT.FR
*IRIT, University of Toulouse III*
*Toulouse, France*

**Stanislav Živný**                                    STANDA.ZIVNY@CS.OX.AC.UK
*Department of Computer Science, University of Oxford*
*Oxford, UK*

## Abstract

The minimisation problem of a sum of unary and pairwise functions of discrete variables is a general NP-hard problem with wide applications such as computing MAP configurations in Markov Random Fields (MRF), minimising Gibbs energy, or solving binary Valued Constraint Satisfaction Problems (VCSPs).

We study the computational complexity of classes of discrete optimisation problems given by allowing only certain types of costs in every *triangle* of variable-value assignments to three distinct variables. We show that for several computational problems, the only non-trivial tractable classes are the well known maximum matching problem and the recently discovered joint-winner property. Our results, apart from giving complete classifications in the studied cases, provide guidance in the search for *hybrid tractable classes*; that is, classes of problems that are not captured by restrictions on the functions (such as submodularity) or the structure of the problem graph (such as bounded treewidth).

Furthermore, we introduce a class of problems with *convex* cardinality functions on *cross-free* sets of assignments. We prove that while imposing only one of the two conditions renders the problem NP-hard, the conjunction of the two gives rise to a novel tractable class satisfying the *cross-free convexity property*, which generalises the joint-winner property to problems of unbounded arity.

## 1. Introduction

The topic of this paper is the following optimisation problem: given a set of discrete variables and a set of functions, each depending on a subset of the variables, minimise the sum of the functions over all variables. This fundamental research problem has been studied within several different contexts of computer science and artificial intelligence under different names: Min-Sum Problems (Werner, 2007), MAP inference in Markov Random Fields (MRF) and Conditional Random Fields (CRF) (Lauritzen, 1996; Wainwright & Jordan, 2008), Gibbs energy minimisation (Geman & Geman, 1984), Valued Constraint Satisfaction Problems (Dechter, 2003), or (for two-state variables) pseudo-Boolean optimisation (Boros & Hammer, 2002).

We use the terminology of Valued Constraint Satisfaction Problems (VCSPs) (Schiex, Fargier, & Verfaillie, 1995; Dechter, 2003). We start with a special case of VCSPs that deals only with the feasibility (rather than optimisation) problem.

A Constraint Satisfaction Problem (CSP) instance consists of a collection of variables which must be assigned values subject to specified constraints (Montanari, 1974). Each CSP instance has an underlying undirected graph, known as its *constraint graph* (or *structure*), whose vertices are the variables of the instance, and two vertices are adjacent if corresponding variables are related by some constraint.

An important line of research on CSPs is to identify all tractable cases which are recognisable in polynomial time. Most of this work has been focused on one of the two general approaches: either identifying forms of constraint which are sufficiently restrictive to ensure tractability no matter how they are combined (Bulatov, Krokhin, & Jeavons, 2005; Feder & Vardi, 1998), or else identifying structural properties of constraint networks which ensure tractability no matter what forms of constraint are imposed (Dechter & Pearl, 1988).

The first approach has led to identifying certain algebraic closure operations known as polymorphisms (Jeavons, 1998) which are necessary for a set of constraint types to ensure tractability. A set of constraint types with this property is called a tractable *constraint language.* The second approach has been used to characterise all tractable cases of bounded-arity CSPs (such as binary CSPs) (Dalmau, Kolaitis, & Vardi, 2002; Grohe, 2007) and unbounded-arity CSPs (Marx, 2010).

In practice, constraint satisfaction problems usually do not possess a sufficiently restricted structure or use a sufficiently restricted constraint language to fall into any of these tractable classes. Nevertheless, they may still have properties which ensure they can be solved efficiently, but these properties concern both the structure and the form of the constraints. Such properties have sometimes been called *hybrid* reasons for tractability (Dechter, 2003; Cohen, 2003; Cohen & Jeavons, 2006; Cooper, Jeavons, & Salamon, 2010; Cohen, Cooper, Green, & Marx, 2011).

CSPs capture only the feasibility aspects of a given problem. Since many computational problems involve seeking a solution that optimises certain criteria, as well as satisfying certain restrictions, various general frameworks for optimisation problems have been studied such as linear programming, mixed integer programming and others (Hooker, 2007). One possibility is to extend CSPs to so-called *soft* constraint satisfaction problems, which allow measures of desirability to be associated with different assignments to the variables (Dechter, 2003; Meseguer, Rossi, & Schiex, 2006). In an instance of a soft CSP, every constraint is associated with a function (rather than a relation as in standard CSPs) which represents preferences among different partial assignments, and the goal is to find the best assignment. Several very general soft CSP frameworks have been proposed in the literature (Schiex, Fargier, & Verfaillie, 1995; Bistarelli, Montanari, & Rossi, 1997). In this paper we focus on one of the very general frameworks, the *valued* constraint satisfaction problem (VCSP) (Schiex, Fargier, & Verfaillie, 1995). VCSPs are powerful enough to include many interesting optimisation problems (Rossi, van Beek, & Walsh, 2006; Cohen, Cooper, Jeavons, & Krokhin, 2006) and, as pointed out at the beginning of this introduction, are equivalent to other well studied optimisation problems studied in computer vision and other fields of computer science and artificial intelligence.

An important line of research on VCSPs is to identify tractable cases which are recognisable in polynomial time. It is well known that structural reasons for tractability generalise to the VCSP (Bertelé & Brioshi, 1972; Dechter, 2003). In the case of language restrictions, only a few conditions are known to guarantee tractability of a given set of valued constraints (Cohen, Cooper, Jeavons, & Krokhin, 2006; Cohen, Cooper, & Jeavons, 2008; Jonsson, Kuivinen, & Thapper, 2011; Kolmogorov, 2011; Kolmogorov & Živný, 2012).

## 1.1 Contributions

This paper is the full version of results described in two conference papers (Cooper & Živný, 2011a, 2011c).

### 1.1.1 Binary VCSPs

In the first part of the paper, we study hybrid tractability of binary VCSPs (i.e. optimisation problems involving functions of at most two arguments) for various sets of possible costs that correspond to CSPs, CSPs with soft unary constraints, Max-CSPs, finite-valued VCSPs and general-valued VCSPs.

We focus on classes of instances defined by allowed combinations of binary costs in every assignment to 3 different variables (called a *triangle*). Our motivation for this investigation is that one such restriction, the so-called joint-winner property has recently been shown to define a tractable class (Cooper & Živný, 2011b). For finite sets of possible costs (corresponding to CSPs and Max-CSPs), there are only finitely many possibilities. For example, in Max-CSPs there are only four possible multi-sets of costs, namely $\{0, 0, 0\}$, $\{0, 0, 1\}$, $\{0, 1, 1\}$ and $\{1, 1, 1\}$. However, for infinite sets of possible costs (corresponding to finite-valued CSPs and general-valued VCSPs) there are infinitely many combinations. Obviously, we cannot consider them all, and hence we consider an equivalence relation based on the total order on the valuation structure. For example, we consider the four equivalence classes of multi-sets $\{\alpha, \beta, \gamma\}$ given by $\alpha = \beta = \gamma$, $\alpha = \beta < \gamma$, $\alpha = \beta > \gamma$, $\alpha < \beta < \gamma$.

For all sets of possible costs $\Omega$ we consider, we prove a dichotomy theorem, thus identifying all tractable cases with respect to the equivalence relation on the combinations of costs. It turns out that there are only two non-trivial tractable cases: the well-known maximum matching problem (Edmonds, 1965b), and the recently discovered joint-winner property (Cooper & Živný, 2011b).

### 1.1.2 Non-binary VCSPs

In the second part of the paper, we introduce the *cross-free convexity property* (CFC), and show that it gives rise to a novel tractable class of VCSPs. Informally speaking, the CFC property is a conjunction of convex cost functions applied to a structured set of sets of variable-value assignments. The CFC property generalises our recent results on VCSPs satisfying the non-overlapping convexity property (Cooper & Živný, 2011b) by dropping the assumption that the input functions are non-decreasing and allowing the assignment-sets to be not only hierarchically nested (laminar) but also cross-free. (All terms will be defined formally in Section 4.) Not only do we generalise the tractable class from the work of Cooper & Živný (2011b), but our algorithm also has better running time compared to the algorithm of Cooper & Živný (2011b). Moreover, we show that relaxing either one of the cross-free or convexity assumptions leads to an NP-hard class.

A VCSP instance may be such that some subset of its constraints are cross-free convex. Since our network is projection-safe (Lee & Leung, 2009), we can use it to establish soft global arc consistency on this subset of constraints viewed as a single global constraint.

We also show that, over Boolean domains, it is possible to determine in polynomial time whether there exists some subset of the constraints such that the VCSP instance satisfies the cross-free convexity property after renaming the variables in these constraints. To explore this area even further, we study restrictions on overlaps of constraint scopes, and identify another tractable class which is incomparable with the cross-free convexity property.

## 1.2 Organisation of the Paper

The rest of this paper is organised as follows. We start, in Section 2, by defining valuation structures, valued constraint satisfaction problems, and basics of flow networks. Section 3 is devoted to the classification of binary VCSPs defined by triangles: In Section 3.1, we present our results on CSPs, followed up with results on CSPs with soft unary constraints in Section 3.2. In Section 3.3, we present our results on Max-CSPs, followed by the results on finite-valued and general-valued VCSPs in Section 3.4 and in Section 3.5 respectively. Section 4 is devoted to our results on non-binary VCSPs: In Section 4.1, we present an algorithm for VCSPs satisfying the cross-free convexity property and analyze its running time. Section 4.4 shows that neither cross-freeness nor convexity on its own is enough to guarantee tractability. In Section 4.5, we extend the class of cross-free convex VCSPs over Boolean domains using the notion of renamability. Section 4.6 explores a related notion over sets of variables rather than sets of variable-value assignments. Finally, we conclude in Section 5.

## 2. Preliminaries

In this section, we define valuation structures, valued constraint satisfaction problems, and present the basics of flow networks.

### 2.1 Valuation Structures

A *valuation structure*, $\Omega$, is a totally ordered set, with a minimum and a maximum element (denoted 0 and $\infty$), together with a commutative, associative binary aggregation operator (denoted $\oplus$), such that for all $\alpha, \beta, \gamma \in \Omega$, $\alpha \oplus 0 = \alpha$, and $\alpha \oplus \gamma \geq \beta \oplus \gamma$ whenever $\alpha \geq \beta$. Members of $\Omega$ are called *costs*.

We shall denote by $\mathbb{Q}_+$ the set of all non-negative rational numbers. We define $\overline{\mathbb{Q}}_+ = \mathbb{Q}_+ \cup \{\infty\}$. We consider the following subsets of the valuation structure $\overline{\mathbb{Q}}_+$: $\{0, \infty\}$, $\{0, 1\}$, $\mathbb{Q}_+$ and $\overline{\mathbb{Q}}_+$, where in all cases the aggregation operation is the standard addition operation on rationals $+$. Moreover, for all $a \in \overline{\mathbb{Q}}_+$, we define $a + \infty = \infty + a = \infty$.

### 2.2 Valued Constraint Satisfaction Problems

An instance of the *Valued Constraint Satisfaction Problem* (VCSP) (Schiex, Fargier, & Verfaillie, 1995) is given by $n$ variables $v_1, \ldots, v_n$ over finite domains $D_1, \ldots, D_n$ of values and a set of constraints $C$. Each constraint from $C$ is a pair $\langle s, g \rangle$, where $s$ is a list of variables $s = \langle v_{i_1}, \ldots, v_{i_m} \rangle$ called the *constraint scope*, and $g$ is an $m$-ary *cost function* $g : D_{i_1} \times \ldots \times D_{i_m} \to \Omega$. Any assignment of values from the domains to all the variables is called a *solution*. The goal is to find an *optimal* solution; that is, a solution which minimises the total cost given by the aggregation of the costs for its restrictions onto each constraint

scope:

$$\min_{v_1 \in D_1, \ldots, v_n \in D_n} \bigoplus_{\langle \langle v_{i_1}, \ldots, v_{i_m} \rangle, g \rangle \in C} g(v_{i_1}, \ldots, v_{i_m}).$$

Depending on the set $\Omega$ of costs which may occur in instances, we get special cases of the VCSP: $\Omega = \{0, \infty\}$ corresponds to the *Constraint Satisfaction Problem* (CSP), $\{0, 1\}$ corresponds to the *Maximum Constraint Satisfaction Problem* (Max-CSP), $\mathbb{Q}_+$ corresponds to the *finite-valued* VCSP, and finally $\overline{\mathbb{Q}}_+$ corresponds to the *general-valued* VCSP.

If all the domains of all the variables are the same, we denote this common domain by $D$. A CSP instance is called *satisfiable* if the cost of an optimal solution is zero (i.e. all constraints are satisfied).

Cost functions with range $\{0, \infty\}$ are called *crisp*. Cost functions which are not crisp are called *soft*.

## 2.3 Binary Valued Constraint Satisfaction Problems

In Section 3 we will be interested in the special case of the VCSP when the bound on the arity of all constraints is 2; these are known as *binary* VCSPs. Without loss of generality, we can assume that any binary VCSP instance contains constraints of all possible scopes; that is, $n$ unary constraints and $\binom{n}{2}$ binary constraints. We denote the cost function associated with the unary constraint with the scope $\langle v_i \rangle$ by $c_i$ and the cost function associated with the binary constraint with the scope $\langle v_i, v_j \rangle$ by $c_{ij}$. The absence of any constraint on variable $v_i$ (or between variables $v_i, v_j$) is modelled by a cost function $c_i$ (or $c_{ij}$, respectively) which is uniformly zero. Using this notation, the goal is to find a solution which minimises the total cost given by:

$$\bigoplus_{i=1}^{n} c_i(v_i) \; \oplus \; \bigoplus_{1 \leq i < j \leq n} c_{ij}(v_i, v_j).$$

**Remark 2.1.** We remark on terminological differences. VCSPs are studied under different names such as Min-Sum, Gibbs energy minimisation, or Markov Random Fields; domain values are sometimes called *labels*, whereas binary instances are called *pairwise* instances, $m$-ary cost functions are called $m$-cliques, and solutions are called *labellings*.

## 2.4 Network Flows

Here we review some basics on flows in graphs. We refer the reader to standard textbooks (Ahuja, Magnanti, & Orlin, 2005; Schrijver, 2003) for more details. We present only the notions and results needed for our purposes. In particular, we deal with integral flows only. We denote by $\mathbb{N}$ the set of positive integers with zero. Let $G = (V, A)$ be a directed graph with vertex set $V$ and arc set $A$. For each arc $a \in A$ there is a *demand/capacity* function $[d(a), c(a)]$ and a *weight* (or cost) function $w(a)$, where $d(a), c(a) \in \mathbb{N}$ and $w(a) \in \mathbb{Q}$. Let $s, t \in V$. A function $f : A \to \mathbb{N}$ is called an $s - t$ *flow* (or just a flow) if for all $v \in V \setminus \{s, t\}$,

$$\sum_{a=(u,v) \in A} f(a) = \sum_{a=(v,u) \in A} f(a) \qquad \text{(flow conservation).}$$

We say that a flow is *feasible* if $d(a) \leq f(a) \leq c(a)$ for each $a \in A$. We define the *value* of flow $f$ as $val(f) = \sum_{a=(s,v)\in A} f(a) - \sum_{a=(v,s)\in A} f(a)$. We define the *cost* of flow $f$ as $\sum_{a \in A} w(a)f(a)$. A *minimum-cost flow* is a feasible flow with minimum cost.

Algorithms for finding the minimum-cost flow of a given value are well known (Ahuja, Magnanti, & Orlin, 2005; Schrijver, 2003). We consider a generalisation of the minimum-cost flow problem. For each arc $a \in A$ there is a convex weight function $w_a$ which associates a cost $w_a(f(a))$ to the flow $f(a)$ along arc $a$. In particular, we consider the model in which the weight functions $w_a$ $(a \in A)$ are convex piecewise linear and given by the breakpoints (which covers the case of convex functions over the integers). The *cost* of flow $f$ is now defined as $\sum_{a \in A} w_a(f(a))$. The corresponding problem of finding a minimum-cost integral flow is known as the *minimum convex cost flow* problem. In a network with $n$ vertices and $m$ edges with capacities at most $U$, the minimum convex cost flow problem can be solved in time $O((m \log U)SP(n, m))$, where $SP(n, m)$ is the time to compute a shortest directed path in a network with $n$ vertices and $m$ edges (Minoux, 1984, 1986; Ahuja, Magnanti, & Orlin, 2005).

## 3. Complexity Classification of Binary VCSPs Defined by Triangles

In a VCSP instance, we use the word *triangle* for any set of assignments $\{\langle v_i, a \rangle, \langle v_j, b \rangle, \langle v_k, c \rangle\}$, where $v_i, v_j, v_k$ are distinct variables and $a \in D_i, b \in D_j, c \in D_k$ are domain values. The multi-set of costs in such a triangle is $\{c_{ij}(a, b), c_{ik}(a, c), c_{jk}(b, c)\}$. A *triple of costs* will always refer to a multi-set of binary costs in a triangle.

A triangle $\{\langle v_i, a \rangle, \langle v_j, b \rangle, \langle v_k, c \rangle\}$, where $a \in D_i, b \in D_j, c \in D_k$, satisfies the *joint-winner property* (JWP) if either all three $c_{ij}(a, b)$, $c_{ik}(a, c)$, $c_{jk}(b, c)$ are the same, or two of them are equal and the third one is bigger. A VCSP instance satisfies the joint-winner property if every triangle satisfies the joint-winner property.

**Theorem 3.1.** (Cooper & Živný, 2011b) *The class of VCSP instances satisfying JWP is tractable.*

In our previous work (Cooper & Živný, 2011b), we also showed that the class defined by the joint-winner property is maximal – allowing a single extra triple of costs that violates the joint-winner property renders the class NP-hard.

**Theorem 3.2.** (Cooper & Živný, 2011b) *Let $\alpha < \beta \leq \gamma$, where $\alpha \in \mathbb{Q}_+$ and $\beta, \gamma \in \overline{\mathbb{Q}}_+$, be a multi-set of costs that do not satisfy the joint-winner property. The class of instances where the costs in each triangle either satisfy the joint-winner property or are $\{\alpha, \beta, \gamma\}$ is NP-hard, even for Boolean Max-CSPs, CSPs over size-3 domains or Boolean finite-valued VCSPs.*

In this section we consider a much broader question, whether allowing any fixed set $S$ of triples of costs in triangles, where $S$ does not necessarily include all triples allowed by the JWP, defines a tractable class of VCSP instances.

In the case of CSP, there are only four possible multi-sets of costs ($\{0, 0, 0\}$, $\{0, 0, \infty\}$, $\{0, \infty, \infty\}$, $\{\infty, \infty, \infty\}$) and it is possible to study all 16 subsets $S$ of this set. But, given an infinite set of possible costs, such as $\mathbb{Q}_+$ or $\overline{\mathbb{Q}}_+$, there is an infinite number of sets $S$ of triples of costs. Obviously, we cannot consider all such sets. Therefore, we only consider cases defined by the total order $<$ on $\Omega$, corresponding to a partition of the set of all possible triples of costs into a small number of types of triples.

Let $\mathfrak{D}$ denote the set of all possible cost types under consideration. Let $\Omega$ be a fixed set of allowed costs. For any $S \subseteq \mathfrak{D}$, we denote by $\mathcal{A}_\Omega(S)$ ($\mathcal{A}$ for allowed) the set of binary VCSP instances whose costs lie in $\Omega$ and where the triples of costs in all triangles belong to $S$.

Our goal is to classify the complexity of $\mathcal{A}_\Omega(S)$ for every $S \subseteq \mathfrak{D}$. The problem $\mathcal{A}_\Omega(S)$ is considered *tractable* if there is a polynomial-time algorithm to solve it and *intractable* if it is NP-hard.

**Proposition 3.3.** *Let $\Omega$ be an arbitrary set of costs and $S$ a set of cost types.*

1. *If $\mathcal{A}_\Omega(S)$ is tractable and $S' \subseteq S$, then $\mathcal{A}_\Omega(S')$ is tractable.*

2. *If $\mathcal{A}_\Omega(S)$ is intractable and $S' \supseteq S$, then $\mathcal{A}_\Omega(S')$ is intractable.*

**Remark 3.4.** We implicitly allow all unary cost functions. In fact, all our tractability results work with unary cost functions, and our NP-hardness results do not require any unary cost functions.

**Remark 3.5.** We consider problems with unbounded domains; that is, the domain sizes are part of the input. However, all our NP-hardness results are obtained for problems with a fixed domain size.[1] In the case of CSPs, we need domains of size 3 to prove NP-hardness, and in all other cases domains of size 2 are sufficient to prove NP-hardness. Since binary CSPs are known to be tractable on Boolean domains, and any VCSP is trivially tractable over domains of size 1, all our NP-hardness results are tight.

## 3.1 CSP

In this section, we will focus on the set of possible costs $\Omega = \{0, \infty\}$; that is, Constraint Satisfaction Problems (CSPs). We consider the four following types of triples of costs:

| Symbol | Costs |
|:------:|:-----:|
| $<$ | $\{0, 0, \infty\}$ |
| $>$ | $\{0, \infty, \infty\}$ |
| $0$ | $\{0, 0, 0\}$ |
| $\infty$ | $\{\infty, \infty, \infty\}$ |

The set of possible cost types is thus $\mathfrak{D} = \{<, >, 0, \infty\}$. Indeed, these four cost types correspond precisely to the four possible multi-sets of costs: $\{0, 0, 0\}$, $\{0, 0, \infty\}$, $\{0, \infty, \infty\}$ and $\{\infty, \infty, \infty\}$. The dichotomy presented in this section therefore represents a complete characterisation of the complexity of CSPs defined by placing restrictions on triples of costs in triangles.

As $\mathcal{A}_{\{0,\infty\}}(\mathfrak{D})$ allows all binary CSPs, $\mathcal{A}_{\{0,\infty\}}(\mathfrak{D})$ is intractable (Papadimitriou, 1994) unless the domain is of size at most 2, in which case it is equivalent to 2-SAT, which is a well-known tractable class (Schaefer, 1978).

---

1. In other words, the considered problems are not fixed-parameter tractable (Downey & Fellows, 1999) in the domain size.

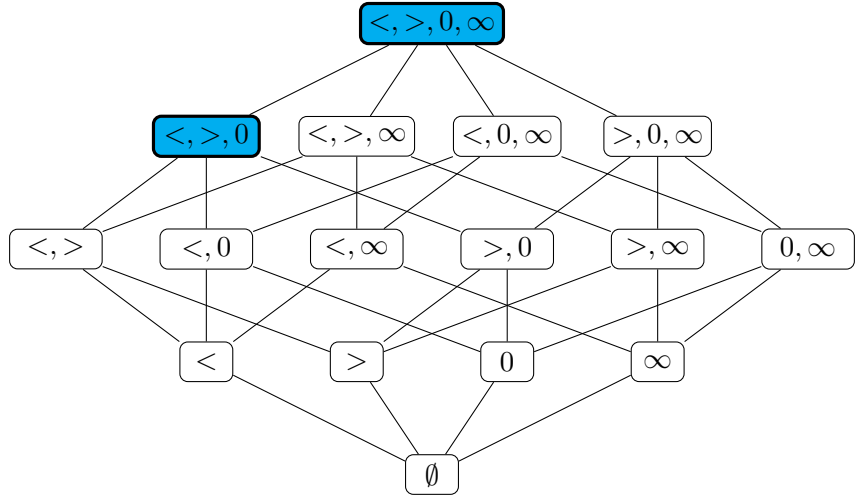

Figure 1: Complexity of CSPs $\mathcal{A}_{\{0,\infty\}}(S), S \subseteq \{<,>,0,\infty\}$.

**Proposition 3.6.** $\mathcal{A}_{\{0,\infty\}}(\mathfrak{D})$ *is intractable unless* $|D| \leq 2$.

The joint-winner property for CSPs gives

**Corollary 3.7** (of Theorem 3.1). $\mathcal{A}_{\{0,\infty\}}(\{<,0,\infty\})$ *is tractable.*

**Proposition 3.8.** $\mathcal{A}_{\{0,\infty\}}(\{>,0,\infty\})$ *is tractable.*

*Proof.* Since $<$ is forbidden, if two binary costs in a triangle are zero then the third binary cost must also be zero. In other words, if the assignment $\langle v_1, a_1 \rangle$ is consistent with $\langle v_i, a_i \rangle$ for each $i \in \{2, \ldots, n\}$, then for all $i, j \in \{1, \ldots, n\}$ such that $i \neq j$, $\langle v_i, a_i \rangle$ is consistent with $\langle v_j, a_j \rangle$. Thus Singleton Arc Consistency, which is a procedure enforcing Arc Consistency for every variable-value pair (Rossi, van Beek, & Walsh, 2006), solves $\mathcal{A}_{\{0,\infty\}}(\{>,0,\infty\})$. ☐

**Proposition 3.9.** $\mathcal{A}_{\{0,\infty\}}(\{<,>,\infty\})$ *is tractable.*

*Proof.* This class is trivial: instances with at least three variables have no solution of finite cost, since the triple of costs $\{0,0,0\}$ is not allowed. ☐

**Proposition 3.10.** $\mathcal{A}_{\{0,\infty\}}(\{<,>,0\})$ *is intractable unless* $|D| \leq 2$.

*Proof.* It is straightforward to encode the 3-Colouring problem as a binary CSP. The result then follows from the fact that 3-Colouring is NP-hard for triangle-free graphs (i.e. graphs that do not contain $K_3$, the complete graph on 3 vertices, as a subgraph), which can be derived from two results from the work of Lovász (1973). (Indeed, 3-Colouring is NP-hard even for triangle-free graphs of degree at most 4; see Maffray & Preissmann, 1996.) The triple of costs $\{\infty, \infty, \infty\}$ cannot occur in the CSP encoding of the colouring of a triangle-free graph. ☐

Results from this section, together with Proposition 3.3, complete the complexity classification, as depicted in Figure 1: white nodes represent tractable cases and shaded nodes represent intractable cases.

**Theorem 3.11.** *For $|D| \geq 3$, a class of binary CSP instances defined as $\mathcal{A}_{\{0,\infty\}}(S)$, where $S \subseteq \{<, >, 0, \infty\}$, is intractable if and only if $\{<, >, 0\} \subseteq S$.*

## 3.2 CSP with Soft Unary Constraints

A simple way to convert classical CSPs into an optimisation problem is to allow soft unary constraints. This framework includes well-studied problems such as Max-Ones over Boolean domains (Creignou, Khanna, & Sudan, 2001; Khanna, Sudan, Trevisan, & Williamson, 2001) and non-Boolean domains (Jonsson, Kuivinen, & Nordh, 2008), Max-Solution (Jonsson & Nordh, 2008), or Min-Cost-Hom (Takhanov, 2010).

It turns out that the dichotomy given in Theorem 3.11 remains valid even if soft unary constraints are allowed. In this case, the intractable cases are now intractable even for domains of size 2.

We use the notation $\mathcal{A}_{\{0,\infty\}}^{\overline{\mathbb{Q}}_+}(S)$ to represent the set of VCSP instances with binary costs from $\{0, \infty\}$, unary costs from $\overline{\mathbb{Q}}_+$ and whose triples of costs in triangles belong to $S$. In other words, we now consider VCSPs with crisp binary constraints and soft unary constraints.

**Theorem 3.12.** *For $|D| \geq 2$, a class of binary CSP instances defined as $\mathcal{A}_{\{0,\infty\}}^{\overline{\mathbb{Q}}_+}(S)$, where $S \subseteq \{<, >, 0, \infty\}$, is intractable if and only if $\{<, >, 0\} \subseteq S$.*

*Proof.* For the tractability part of the theorem, it suffices to show tractability when $S$ is $\{<, >, \infty\}$, $\{<, 0, \infty\}$ or $\{>, 0, \infty\}$, the three maximal tractable sets in the case of CSP shown in Figure 1.

The tractability of $\mathcal{A}_{\{0,\infty\}}^{\overline{\mathbb{Q}}_+}(\{<, 0, \infty\})$ is again a corollary of Theorem 3.1 since the joint-winner property allows any unary soft constraints.

To solve $\mathcal{A}_{\{0,\infty\}}^{\overline{\mathbb{Q}}_+}(\{>, 0, \infty\})$ in polynomial time, we establish Singleton Arc Consistency in the CSP instance corresponding to the binary constraints and then loop over all assignments to the first variable. For each assignment $a_1$ to variable $v_1$, we can determine the optimal global assignment which is an extension of $\langle v_1, a_1 \rangle$ by simply choosing the assignment $a_i$ for each variable $v_i$ with the least unary cost $c_i(a_i)$ among those assignments $\langle v_i, a_i \rangle$ that are consistent with $\langle v_1, a_1 \rangle$.

As in the proof of Proposition 3.9, any instance of $\mathcal{A}_{\{0,\infty\}}^{\overline{\mathbb{Q}}_+}(\{<, >, \infty\})$ is tractable, since instances with at least three variables have no solution of finite cost.

Sets $S$ which are intractable for CSPs clearly remain intractable when soft unary constraints are allowed. However, we want to prove intractability even in the Boolean case; that is, when $|D| = 2$.

The intractability of $\mathcal{A}_{\{0,\infty\}}^{\overline{\mathbb{Q}}_+}(\{<, >, 0\})$ (and hence, by Proposition 3.3, of $\mathcal{A}_{\{0,\infty\}}^{\overline{\mathbb{Q}}_+}(\{<, >, 0, \infty\})$) follows from the fact that the Independent Set problem (Garey & Johnson, 1979) is intractable even on triangle free graphs. This follows from the standard trick (Poljak, 1974) of replacing every edge by $P_4$, the path on 4 vertices (this operation is also known as 2-subdivision). In particular, a graph $G$ with $m$ edges has an independent set of size $k$ if and only if the 2-subdivision of $G$, denoted by $G'$, has an independent set of size $k + m$. Note that $G'$ is triangle-free. Any instance $G'$ of the Independent Set problem on triangle-free graphs can be encoded as an instance of $\mathcal{A}_{\{0,\infty\}}^{\overline{\mathbb{Q}_+}}(\{<,>,0\})$ over the $\{0,1\}$ domain in the straightforward way: variables correspond to vertices; edge $\{i,j\}$ yields cost function $c_{ij}(1,1) = \infty$ and $c_{ij}(x,y) = 0$ for $(x,y) \neq (1,1)$; $c_i(0) = 1$ and $c_i(1) = 0$ for every $i$. Since $G'$ is triangle-free, the constructed instance belongs to $\mathcal{A}_{\{0,\infty\}}^{\overline{\mathbb{Q}_+}}(\{<,>,0\})$. $\qquad\square$

### 3.3 Max-CSP

In this section, we will focus on the set of possible costs $\Omega = \{0,1\}$. It is well known that the VCSP with costs in $\{0,1\}$ is polynomial-time equivalent to unweighted Max-CSP (no repetition of constraints allowed) (Rossi, van Beek, & Walsh, 2006). The four types of triples of costs we consider are:

| Symbol | Costs |
|:------:|:------:|
| $<$ | $\{0,0,1\}$ |
| $>$ | $\{0,1,1\}$ |
| $0$ | $\{0,0,0\}$ |
| $1$ | $\{1,1,1\}$ |

The set of possible cost types is then $\mathfrak{D} = \{<,>,0,1\}$. Again, these four costs types correspond precisely to the four possible multi-sets of costs: $\{0,0,0\}$, $\{0,0,1\}$, $\{0,1,1\}$, and $\{1,1,1\}$. As for the CSP, our dichotomy result for Max-CSP represents a complete characterisation of the complexity of classes of instances defined by placing restrictions on triples of costs in triangles.

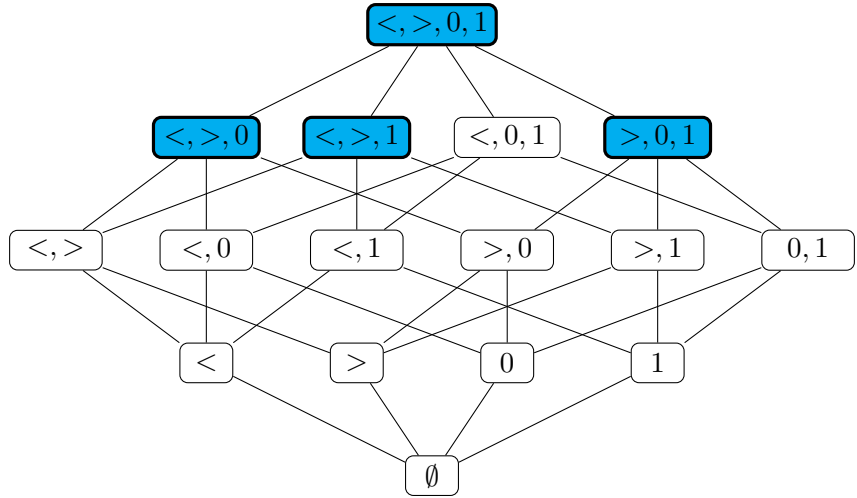

Figure 2: Complexity of Max-CSPs $\mathcal{A}_{\{0,1\}}(S), S \subseteq \{<,>,0,1\}$.

As $\mathcal{A}_{\{0,1\}}(\mathfrak{D})$ allows all binary Max-CSPs, $\mathcal{A}_{\{0,1\}}(\mathfrak{D})$ is intractable (Garey & Johnson, 1979; Papadimitriou, 1994) unless the domain is of size 1.

**Proposition 3.13.** $\mathcal{A}_{\{0,1\}}(\mathfrak{D})$ *is intractable unless* $|D| \leq 1$.

The joint-winner property (Cooper & Živný, 2011b) for Max-CSPs gives

**Corollary 3.14** (of Theorem 3.1). $\mathcal{A}_{\{0,1\}}(\{<, 0, 1\})$ *is tractable.*

**Proposition 3.15.** $\mathcal{A}_{\{0,1\}}(\{<, >\})$ *is tractable.*

*Proof.* We show that $\mathcal{A}_{\{0,1\}}(\{<, >\})$ contains instances on at most 5 variables, thus showing that $\mathcal{A}_{\{0,1\}}(\{<, >\})$ is trivially tractable. Consider an instance of $\mathcal{A}_{\{0,1\}}(\{<, >\})$ on 6 or more variables. Choose 6 arbitrary variables $v_1, \ldots, v_6$ and 6 domain values $d_i \in D_{v_i}$, $1 \leq i \leq 6$. Every cost is either 0 or 1. It is well known (Goodman, 1959) and not difficult to show[2] that for every 2-colouring of edges of $K_6$ (the complete graph on 6 vertices) there is a monochromatic triangle. Therefore, there is a triangle with costs either $\{0, 0, 0\}$ or $\{1, 1, 1\}$. But this is a contradiction with the fact that only cost types $<$ (i.e. $\{0, 0, 1\}$) and $>$ (i.e. $\{1, 1, 0\}$) are allowed. $\qquad \square$

**Remark 3.16.** Both $\mathcal{A}_{\Omega}(\{>\})$ and $\mathcal{A}_{\Omega}(\{<, >\})$ are tractable over any finite set of costs $\Omega$ due to a similar Ramsey type of argument: given $\Omega = \{0, 1, \ldots, K - 1\}$, there is $n_K \in \mathbb{N}$ such that for every complete graph $G$ on $n$ vertices, where $n \geq n_K$, and every colouring of the edges of $G$ with $K$ colours, there is a monochromatic triangle in $G$. Hence there are only finitely many instances, which can be stored in a look-up table. However, once the set of costs is infinite (e.g. $\mathbb{Q}_+$), both classes become intractable, as shown in the next section.

**Proposition 3.17.** $\mathcal{A}_{\{0,1\}}(\{>, 0, 1\})$ *is intractable unless* $|D| \leq 1$.

*Proof.* Given an instance of the Max-2SAT problem, we show how to reduce it to a $\{0, 1\}$-valued VCSP instance from $\mathcal{A}_{\{0,1\}}(\{>, 0, 1\})$. The result then follows from the well-known fact that Max-2SAT is NP-hard (Garey & Johnson, 1979; Papadimitriou, 1994). Recall that an instance of Max-2SAT is given by a set of $m$ clauses of length 2 over $n$ variables $x_1, \ldots, x_n$ and the goal is to find an assignment that maximises the number of clauses that have at least one true literal.

In order to simplify notation, rather than constructing a VCSP instance from $\mathcal{A}_{\{0,1\}}(\{>, 0, 1\})$ with the goal to minimise the total cost, we construct an instance from $\mathcal{A}_{\{0,1\}}(\{<, 0, 1\})$ with the goal to maximise the total cost. This implies that the allowed sets of costs in triangles are $\{0, 0, 1\}$, $\{0, 0, 0\}$, and $\{1, 1, 1\}$. Clearly, these two problems are polynomial-time equivalent.

For each variable $x_i$, we create a large number $M$ of copies $x_i^j$ of $x_i$ with domain $\{0, 1\}$, $1 \leq i \leq n$ and $1 \leq j \leq M$. For each variable $x_i$, the new copies of $x_i$ are pairwise joined by an equality-encouraging cost function $h$, where $h(x, y) = 1$ if $x = y$ and $h(x, y) = 0$ otherwise. By choosing $M$ very large, we can assume from now on that all copies of $x_i$ will be assigned the same value in all optimal solutions. We can effectively ignore the contribution of these cost functions, which is $K = n\binom{M}{2}$, to the total cost. It is straightforward to check that all triangles involving the new copies of the variables have the allowed costs.

---

2. Take an arbitrary vertex $v$ in $K_6$ where every edge is coloured either blue or red. By the pigeonhole principle, $v$ is incident to at least 3 blue or at least 3 red edges. Without loss of generality, we consider the former case. Let $v_1, v_2$ and $v_3$ be the three vertices incident to three blue edges incident to $v$. If an any of the edges $\{v_1, v_2\}$, $\{v_1, v_3\}$, $\{v_2, v_3\}$ is blue, we have a blue triangle. If all three edges are red, we have a red triangle.

For each clause $(l_1 \vee l_2)$, where $l_1$ and $l_2$ are literals, we create a variable $z_i$ with domain $\{l_1, l_2\}$, $1 \leq i \leq m$. For each literal $l$ in the domain of $z_k$: if $l$ is a positive literal $l = x_i$, we introduce cost function $g$ between $z_k$ and each copy $x_i^j$ of $x_i$, where $g(l, 1) = 1$ and $g(., .) = 0$ otherwise; if $l$ is a negative literal $l = \neg x_i$, we introduce cost function $g'$ between $z_k$ and each copy $x_i^j$ of $x_i$, where $g'(l, 0) = 1$ and $g'(., .) = 0$ otherwise.

To make sure that the only sets of costs in all triangles are $\{0, 0, 1\}$, $\{0, 0, 0\}$, and $\{1, 1, 1\}$, we also add cost functions $f$ between the different clause variables $z_k$ and $z_{k'}$ involving the same literal $l$, where $f(l, l) = 1$ and $f(., .) = 0$ otherwise. The contribution of all the cost functions between $z_k$ and $z_{k'}$, $1 \leq k \neq k' \leq m$, is less than $M$ and hence of no importance for $M$ very large.

Answering the question of whether the resulting VCSP instance has a solution with a cost $\geq K + pM$ is equivalent to determining whether the original Max-2SAT instance has a solution satisfying at least $p$ clauses. This is because each clause variable $z_k$ can only add a score $\geq M$ if we assign value $l$ to $z_k$ for some literal $l$ which is assigned true. $\qquad \square$

**Proposition 3.18.** *Both $\mathcal{A}_{\{0,1\}}(\{<, >, 0\})$ and $\mathcal{A}_{\{0,1\}}(\{<, >, 1\})$ are intractable unless $|D| \leq 1$.*

*Proof.* We present a reduction from MAX-CUT, a well-known NP-hard problem (Garey & Johnson, 1979), which is NP-hard even on triangle-free graphs (Lewis & Yannakakis, 1980). An instance of MAX-CUT can easily be modelled as a Boolean $\{0, 1\}$-valued VCSP instance: every vertex of the graph is represented by a variable with the Boolean domain $\{0, 1\}$, and every edge yields cost function $f$, where $f(x, y) = 1$ if $x = y$ and $f(x, y) = 0$ if $x \neq y$. Observe that since the original graph is triangle-free, there cannot be a triangle with costs $\{1, 1, 1\}$. Therefore, the constructed instance belongs to $\mathcal{A}_{\{0,1\}}(\{<, >, 0\})$.

For the $\mathcal{A}_{\{0,1\}}(\{<, >, 1\})$ case, instead of minimising the total cost, we maximise the total cost for instances from $\mathcal{A}_{\{0,1\}}(\{<, >, 0\})$. Again, we model an instance of the MAX-CUT problem using Boolean variables, and every edge yields a cost function $g$, where $g(x, y) = 0$ if $x = y$ and $g(x, y) = 1$ if $x \neq y$ (where in this case the aim is to maximise the total cost). The constructed instance belongs to $\mathcal{A}_{\{0,1\}}(\{<, >, 0\})$. (In fact, in this case we do not need the original graph to be triangle-free.) $\qquad \square$

**Proposition 3.19.** *$\mathcal{A}_{\{0,1\}}(\{>, 0\})$ is tractable.*

*Proof.* Let $I$ be an instance from $\mathcal{A}_{\{0,1\}}(\{>, 0\})$. The algorithm loops through all possible assignments $\{\langle v_1, a_1 \rangle, \langle v_2, a_2 \rangle\}$ to the first two variables. Suppose that $c_{12}(a_1, a_2) = 1$ (the case $c_{12}(a_1, a_2) = 0$ is similar). Observe that the possible variable-value assignments to other variables $\{\langle v_i, b \rangle \,|\, 3 \leq i \leq n, b \in D_i\}$ can be uniquely split in two sets $L$ and $R$ such that: (1) for every $\langle v_i, b \rangle \in L$, $c_{1i}(a_1, b) = 1$ and $c_{2i}(a_2, b) = 0$; for every $\langle v_i, b \rangle, \langle v_j, c \rangle \in L$, $c_{ij}(b, c) = 0$; (2) for every $\langle v_i, b \rangle \in R$, $c_{1i}(a_1, b) = 0$ and $c_{2i}(a_2, b) = 1$; for every $\langle v_i, b \rangle, \langle v_j, c \rangle \in R$, $c_{ij}(b, c) = 0$; (3) for every $\langle v_i, b \rangle \in L$ and $\langle v_j, c \rangle \in R$, $c_{ij}(b, c) = 1$. Ignoring unary cost functions for a moment, to find an optimal assignment to the remaining $n - 2$ variables, one has to decide how many variables $v_i$, $3 \leq i \leq n$, will be assigned a value $b \in D_i$ such that $\langle v_i, b \rangle \in L$. The cost of a global assignment involving $k$ variable-value assignments from $L$ is $1 + k + (n - 2 - k) + k(n - 2 - k) = n - 1 + k(n - 2 - k)$. For some variables $v_i$ it could happen that $\langle v_i, b \rangle \in L$ for all $b \in D_i$ or $\langle v_i, c \rangle \in R$ for all $c \in D_i$. If this is the case, then we choose an arbitrary value $b$ for $x_i$ with minimum unary cost $c_i(b)$. This is an optimal choice whatever the assignments to the variables $x_j$ ($j \in \{3, \ldots, i-1, i+1, \ldots, n\}$).

Assuming that all such variables have been eliminated and now taking into account unary cost functions, the function to minimise is given by the objective function (in which we drop the constant term $n - 1$):

$$\left( \sum x_i \right)(n - 2 - \sum x_i) \; + \; \sum w_i^L x_i \; + \; \sum w_i^R (1 - x_i)$$

(each sum being over $i \in \{3, \ldots, n\}$), where $x_i \in \{0, 1\}$ indicates whether $v_i$ is assigned a value from $R$ or $L$, $w_i^L = \min\{c_i(b) : b \in D_i \wedge \langle v_i, b \rangle \in L\}$, and similarly $w_i^R = \min\{c_i(c) : c \in D_i \wedge \langle v_i, c \rangle \in R\}$. The objective function is thus equal to $k(n-2-k) + \sum w_i^L x_i + \sum w_i^R(1-x_i)$, where, as above, $k = \sum x_i$ is the number of assignments from $L$. This objective function is minimised either when $k = 0$ or when $k = n - 2$. This follows from the fact that the contribution of unary cost functions to the objective function is $\sum w_i^L x_i + \sum w_i^R (1 - x_i)$ which is at most $n - 2$ (since in Max-CSP all unary costs belong to $\{0, 1\}$). This is no greater than the value of the quadratic term $k(n-2-k)$ for all values of $k$ in $\{1, \ldots, n-3\}$, i.e. not equal to 0 or $n - 2$.

The optimal assignment which involves $k = 0$ (respectively $k = n - 2$) assignments from $L$ is obtained by simply choosing each value $a_i$ (for $i > 2$) with minimum unary cost among all assignments $\langle v_i, a_i \rangle \in R$ (respectively $L$).

In the case that $c_{12}(a_1, a_2) = 0$, a similar argument shows that the quadratic term in the objective function is now $2(n-2-k) + k(n-2-k) = (k+2)(n-2-k)$. This is always minimised by setting $k = n - 2$ and again the sum of the unary costs is no greater than the value of the quadratic term for other values of $k \neq n - 2$. The optimal assignment which involves all $k = n - 2$ assignments from $L$ is obtained by simply choosing each value $a_i$ (for $i > 2$) with minimum unary cost among all assignments $\langle v_i, a_i \rangle \in L$. $\qquad \square$

**Proposition 3.20.** $\mathcal{A}_{\{0,1\}}(\{>, 1\})$ *is tractable.*

*Proof.* Let $I$ be an instance from $\mathcal{A}_{\{0,1\}}(\{>, 1\})$ without any unary constraints; i.e. all constraints are binary. Observe that every variable-value assignment $\langle v_i, a \rangle$, where $a \in D_i$, is included in zero-cost assignment-pairs involving at most one other variable; i.e. there is at most one variable $v_j$, such that $c_{ij}(a, b) = 0$ for some $b \in D_j$. In order to minimise the total cost, we have to maximise the number of zero-cost assignment-pairs. In a global assignment, no two zero-cost assignment-pairs can involve the same variable, which means that this can be achieved by a reduction to the maximum matching problem, a problem solvable in polynomial time (Edmonds, 1965b). We build a graph with vertices given by the variables of $I$, and there is an edge $\{v_i, v_j\}$ if and only if there is $a \in D_i$ and $b \in D_j$ such that $c_{ij}(a, b) = 0$.

To complete the proof, we show that unary constraints do not make the problem more difficult to solve; it suffices to perform a preprocessing step before the reduction to maximum matching. Let $v_i$ be an arbitrary variable of $I$. If $c_i(a) = 1$ for all $a \in D_i$, then we can effectively ignore the unary cost function $c_i$ since it simply adds a cost of 1 to any solution. Otherwise, we show that all $a \in D_i$ such that $c_i(a) = 1$ can be ignored. Take an arbitrary assignment $s$ to all variables such that $s(v_i) = a$, where $c_i(a) = 1$. Now take any $b \in D_i$ such that $c_i(b) = 0$. We claim that assignment $s'$ defined by $s'(v_i) = b$ and $s'(v_j) = s(v_j)$ for every $j \neq i$ does not increase the total cost compared with $s$. Since the assignment $\langle v_i, a \rangle$ can occur in at most one zero-cost assignment-pair, there are two cases to consider: (1) if there is no $\langle v_j, c \rangle$ with $s(v_j) = c$ such that $c_{ij}(a, c) = 0$, then the claim holds since $c_i(a) = 1$ and $c_i(b) = 0$, so the overall cost can only decrease if we replace $a$ by $b$; (2) if there is exactly one $j \neq i$ such that $c_{ij}(a, c) = 0$ and $s(v_j) = c$, then again the cost of $s'$ cannot increase because the possible increase of cost by 1 in assigning $b$ to $v_i$ is compensated by the unary cost function $c_i$. Therefore, before using the reduction to maximum matching, we can remove all $a \in D_i$ such that $c_i(a) = 1$ and keep only those $a \in D_i$ such that $c_i(a) = 0$. □

**Remark 3.21.** In the proof of Proposition 3.20, we have shown that any instance from $\mathcal{A}_{\{0,1\}}(\{>,1\})$ can be reduced to an instance of maximum matching in graphs (Edmonds, 1965b). We remark that conversely, given a graph $G$, the maximum matching problem in $G$ can be modelled as a VCSP instance $I'_G$ from $\mathcal{A}_{\{0,1\}}(\{>,1\})$.

We order the vertices of $G$ arbitrarily and call them $1, 2, \ldots, n$. The instance $I_G$ will have $n$ variables $v_1, \ldots, v_n$, one for each vertex of $G$. Let $\{n_1, \ldots, n_m\}$ be the neighbours of vertex $i$ in $G$, where $m$ is the degree of vertex $i$ in $G$; that is, $\{j \mid \{i,j\} \in E(G)\} = \{n_1, \ldots, n_m\}$. We define $D_i = \{0, n_1, \ldots, n_m\}$.

Any edge $\{i, j\} \in E(G)$, where $i < j$, yields $c_{ij}(j, i) = 1$, and all remaining costs are 0. It follows from the definition of $I_G$ that (i) solutions to $I_G$ of maximum cost correspond to maximum matchings in $G$; and (ii) $I_G \in \mathcal{A}_{\{0,1\}}(\{<, 0\})$. By swapping the costs 0 and 1, we get an instance $I'_G$ from $\mathcal{A}_{\{0,1\}}(\{>, 1\})$, whose solutions correspond to matchings in $G$ and solutions of minimum cost correspond to maximum matchings in $G$.

Results from this section, together with Proposition 3.3, complete the complexity classification, as depicted in Figure 2: white nodes represent tractable cases and shaded nodes represent intractable cases.

**Theorem 3.22.** *For* $|D| \geq 2$, *a class of binary unweighted Max-CSP instances defined as* $\mathcal{A}_{\{0,1\}}(S)$, *where* $S \subseteq \{<, >, 0, 1\}$, *is intractable if and only if either* $\{<, >, 0\} \subseteq S$, $\{<, >, 1\} \subseteq S$, *or* $\{>, 0, 1\} \subseteq S$.

## 3.4 Finite-Valued VCSP

In this section, we will focus on finite-valued VCSPs. In other words, we consider the set of possible costs $\Omega = \mathbb{Q}_+$. Since there are an infinite number of triples of costs, we consider types of triples defined by the total order on $\Omega$. We study three different ways of partitioning the set of all triples of costs into distinct types.

### 3.4.1 Classification with respect to Order

The set of possible cost types is $\mathfrak{D} = \{\triangle, <, >, =\}$, where these four types are defined in the following table:

| Symbol | Costs | Remark |
|:------:|:-----:|:------:|
| $\triangle$ | $\{\alpha, \beta, \gamma\}$ | $\alpha, \beta, \gamma \in \Omega,\ \alpha \neq \beta \neq \gamma \neq \alpha$ |
| $<$ | $\{\alpha, \alpha, \beta\}$ | $\alpha, \beta \in \Omega,\ \alpha < \beta$ |
| $>$ | $\{\alpha, \alpha, \beta\}$ | $\alpha, \beta \in \Omega,\ \alpha > \beta$ |
| $=$ | $\{\alpha, \alpha, \alpha\}$ | $\alpha \in \Omega$ |

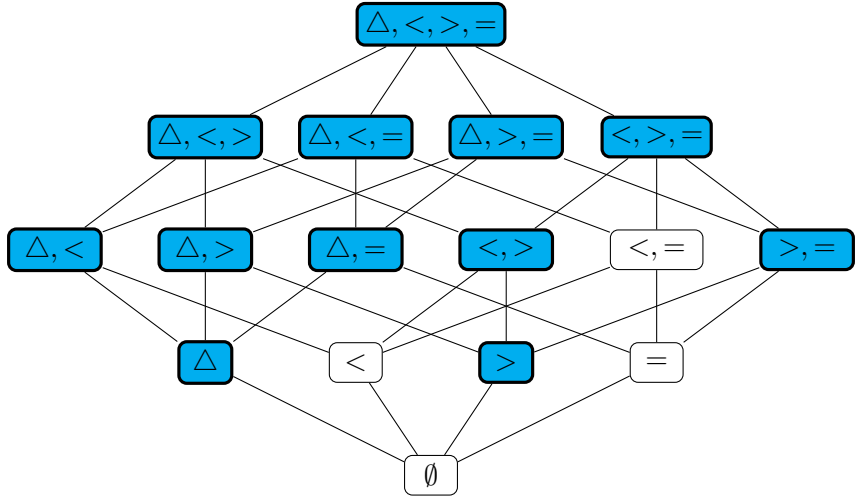

Figure 3: Complexity of finite-valued VCSPs $\mathcal{A}_{\mathbb{Q}_+}(S), S \subseteq \{\triangle, <, >, =\}$.

As $\mathcal{A}_{\mathbb{Q}_+}(\mathfrak{D})$ allows all finite-valued VCSPs, it is intractable even over a Boolean domain (Cohen, Cooper, Jeavons, & Krokhin, 2006) as it includes the Max-SAT problem for the exclusive or predicate (Papadimitriou & Yannakakis, 1991; Creignou, Khanna, & Sudan, 2001).

**Proposition 3.23.** $\mathcal{A}_{\mathbb{Q}_+}(\mathfrak{D})$ *is intractable unless* $|D| \leq 1$.

The joint-winner property (Cooper & Živný, 2011b) for finite-valued VCSPs gives

**Corollary 3.24** (of Theorem 3.1). $\mathcal{A}_{\mathbb{Q}_+}(\{<, =\})$ *is tractable.*

**Proposition 3.25.** $\mathcal{A}_{\mathbb{Q}_+}(\{\triangle\})$ *is intractable unless* $|D| \leq 1$.

*Proof.* We show a reduction from Max-Cut, a well-known NP-hard problem (Garey & Johnson, 1979). An instance of Max-Cut can be easily modelled as a Boolean finite-valued VCSP instance: every vertex of the graph is represented by a variable with the Boolean domain $\{0, 1\}$, and every edge yields cost function $f$, where $f(x, y) = 1$ if $x = y$ and $f(x, y) = 0$ if $x \neq y$. However, the constructed instance does not belong to $\mathcal{A}_{\mathbb{Q}_+}(\{\triangle\})$. Nevertheless, we can amend the VCSP instance by infinitesimal perturbations: all occurrences of the cost 0 are replaced by different numbers that are very close to 0, and all occurrences of the cost 1 are replaced by different numbers very close to 1. Now since all the costs are different, clearly the instance belongs to $\mathcal{A}_{\mathbb{Q}_+}(\{\triangle\})$. □

**Proposition 3.26.** $\mathcal{A}_{\mathbb{Q}_+}(\{>\})$ *is intractable unless* $|D| \leq 1$.

*Proof.* We prove this by a perturbation of the construction in the proof of Proposition 3.17, which shows intractability of $\mathcal{A}_{\mathbb{Q}_+}(\{>, =\})$. In order to simplify the proof, similarly to the proof of Proposition 3.17, we prove that maximising the total cost in the class $\mathcal{A}_{\mathbb{Q}_+}(\{<\})$ is NP-hard.

In the construction in the proof of Proposition 3.17 we add $i\epsilon$ to each binary cost $c_{ij}(a, b)$, where $i < j$, if $c_{ij}(a, b)$ was equal to 1. We assume that $\epsilon$ is very small ($n\epsilon < 1$). This simply ensures that each triple of costs $\{1, 1, 1\}$ in a triangle of assignments is now perturbed to become $\{1 + i\epsilon, 1 + i\epsilon, 1 + j\epsilon\}$.

In the reduction from Max-2SAT, for each literal $l$, let $C_l$ be the set of all variable-value assignments corresponding to $l$ (in both the $x_i^j$ and the $z_k$ variables). Recall that all binary costs for pairs of the assignments within $C_l$ were 1 and all binary costs for pairs of the assignments from distinct $C_l, C_{l'}$ were all 0 in the VCSP encoding of the Max-2SAT instance. We place an arbitrary ordering on the literals $l_1 < l_2 < \cdots < l_r$. We then add $i\epsilon$ to each binary cost between two variable-value assignments whenever these assignments correspond to literals $l_i, l_j$ with $i < j$. This simply ensures that each triple of costs $\{0, 0, 0\}$ in a triangle of assignments is now perturbed to become $\{0 + i\epsilon, 0 + i\epsilon, 0 + j\epsilon\}$.

The resulting VCSP instance is in $\mathcal{A}_{\mathbb{Q}_+}(\{>\})$ and correctly codes the original Max-2SAT instance for sufficiently small $\epsilon$. $\qquad\square$

Results from this section, together with Proposition 3.3, complete the complexity classification, as depicted in Figure 3: white nodes represent tractable cases and shaded nodes represent intractable cases.

**Theorem 3.27.** *For $|D| \geq 2$, a class of binary finite-valued VCSP instances defined as $\mathcal{A}_{\mathbb{Q}_+}(S)$, where $S \subseteq \{\triangle, <, >, =\}$, is tractable if and only if $S \subseteq \{<, =\}$.*

### 3.4.2 CLASSIFICATION WITH RESPECT TO MINIMUM COST

The tractable classes $\mathcal{A}_{\{0,1\}}(\{>, 1\})$, $\mathcal{A}_{\{0,1\}}(\{>, 0\})$ and $\mathcal{A}_{\{0,1\}}(\{<, >\})$ appear in Figure 2, but do not appear as subclasses of the tractable classes $\mathcal{A}_{\mathbb{Q}_+}(S)$ identified in Figure 3. This is due to the fact that for the infinite set of possible costs $\Omega = \mathbb{Q}_+$, Figure 3 covers only a subset of the infinite number of possible restrictions on triples of costs in triangles. We now consider triples of costs which allow us to find generalisations of these three tractable classes to finite-valued VCSPs, by considering restrictions depending on the relationship of costs with the minimum or maximum binary cost in an instance.

We start with the minimum cost. Without loss of generality we can assume that the minimum binary cost of an instance is 0. We consider the following types of triples of costs:

| Symbol | Costs | Remark |
|:------:|:-----:|:------:|
| $\triangle^0$ | $\{\alpha, \beta, 0\}$ | $\alpha, \beta \in \Omega, \; \alpha > \beta > 0$ |
| $<^0$ | $\{0, 0, \alpha\}$ | $\alpha \in \Omega, \; \alpha > 0$ |
| $>^0$ | $\{\alpha, \alpha, 0\}$ | $\alpha \in \Omega, \; \alpha > 0$ |
| $0$ | $\{0, 0, 0\}$ | |

For simplicity of presentation, we do not consider the remaining type of triples of costs, namely $\{\alpha, \beta, \gamma\}$ such that $\alpha, \beta, \gamma > 0$. Since it is possible to transform any VCSP instance into an equivalent instance with non-zero costs by adding a constant $\epsilon > 0$ to all binary costs, it is clear that allowing all such triples of costs would render the VCSP intractable.

The complexity of combinations of costs from $\{\triangle^0, <^0, >^0, 0\}$ are shown in Figure 4: white nodes represent tractable cases and shaded nodes represent intractable cases.

**Proposition 3.28.** $\mathcal{A}_{\mathbb{Q}_+}(\{>^0, 0\})$ *is tractable.*

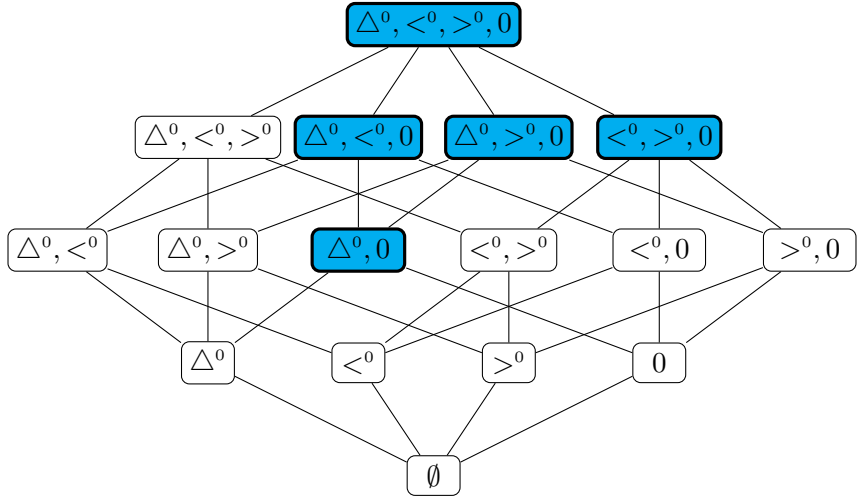

Figure 4: Complexity of finite-valued VCSPs $\mathcal{A}_{\mathbb{Q}_+}(S)$, $S \subseteq \{\triangle^0, <^0, >^0, 0\}$.

*Proof.* Observe that either all non-zero binary costs involve the same variable $v_k$ (i.e. $c_{ij} = 0$ for all $i, j \neq k$) or there is only one distinct cost $\alpha > 0$ in the instance. (Otherwise, if there are two distinct $\alpha \neq \beta$ non-zero costs $\alpha, \beta > 0$ in the instance such that $c_{ij}(a,b) = \alpha$ and $c_{kl}(c,d) = \beta$ for distinct $i, j, k, l$, then it is easy to verify that it is not possible to assign costs to $c_{ik}(a,c)$, $c_{il}(a,d)$, $c_{jk}(b,c)$, $c_{jl}(b,d)$ so that all triangles have cost types $>^0$ or 0.) This implies that $\mathcal{A}_{\mathbb{Q}_+}(\{>^0, 0\})$ is equivalent to $\mathcal{A}_{\{0,1\}}(\{>, 0\})$ after the instantiation of at most one variable. $\square$

**Corollary 3.29** (of Theorem 3.1). $\mathcal{A}_{\mathbb{Q}_+}(\{<^0, 0\})$ *is tractable.*

**Proposition 3.30.** $\mathcal{A}_{\mathbb{Q}_+}(\{\triangle^0, <^0, >^0\})$ *is tractable.*

*Proof.* Analogously to the Ramsey type argument in the proof of Proposition 3.15, any instance on more than 5 variables must contain either a triangle of zero costs or a triangle of three non-zero costs and hence cannot belong to $\mathcal{A}_{\mathbb{Q}_+}(\{\triangle^0, <^0, >^0\})$. $\square$

**Proposition 3.31.** $\mathcal{A}_{\mathbb{Q}_+}(\{<^0, >^0, 0\})$ *is intractable unless* $|D| \leq 1$.

*Proof.* By reduction from MAX-CUT on triangle-free graphs as in the proof of Proposition 3.18 $\square$

**Proposition 3.32.** $\mathcal{A}_{\mathbb{Q}_+}(\{\triangle^0, 0\})$ *is intractable unless* $|D| \leq 1$.

*Proof.* It has been shown that the VCSP remains intractable on bipartite graphs and Boolean domains (Cooper & Živný, 2011b). Let $I$ be such an instance with a partition $V_1, V_2$ of the variables. Insignificantly small but distinct costs can be added to all binary costs in $I$ between variables $i \in V_1$ and $j \in V_2$ to ensure that all triangles are of type $\triangle^0$ or 0. $\square$

**Theorem 3.33.** *For $|D| \geq 2$, a class of binary finite-valued VCSP instances defined as $\mathcal{A}_{\mathbb{Q}_+}(S)$, where $S \subseteq \{\triangle^0, <^0, >^0, 0\}$, is tractable if and only if $S \subseteq \{<^0, 0\}$, $S \subseteq \{>^0, 0\}$ or $S \subseteq \{\triangle^0, <^0, >^0\}$.*

### 3.4.3 CLASSIFICATION WITH RESPECT TO MAXIMUM COST

Let $M \in \mathbb{Q}_+$ be any cost and consider the following types of triples of costs:

| Symbol | Costs | Remark |
|:---:|:---:|:---:|
| $\triangle^M$ | $\{\alpha, \beta, M\}$ | $\alpha, \beta \in \Omega, \; \alpha < \beta < M$ |
| $<^M$ | $\{\alpha, \alpha, M\}$ | $\alpha \in \Omega, \; \alpha < M$ |
| $>^M$ | $\{\alpha, M, M\}$ | $\alpha \in \Omega, \; \alpha < M$ |
| $M$ | $\{M, M, M\}$ | |

Again, we do not consider the remaining type of triples of costs, namely $\{\alpha, \beta, \gamma\}$ such that $\alpha, \beta, \gamma < M$, since allowing such triples of costs renders the VCSP intractable. If $\{\triangle^M, <^M, >^M, M\}$ are the only allowed combinations of triples of costs, then $M$ is clearly the maximum binary cost in the instance.

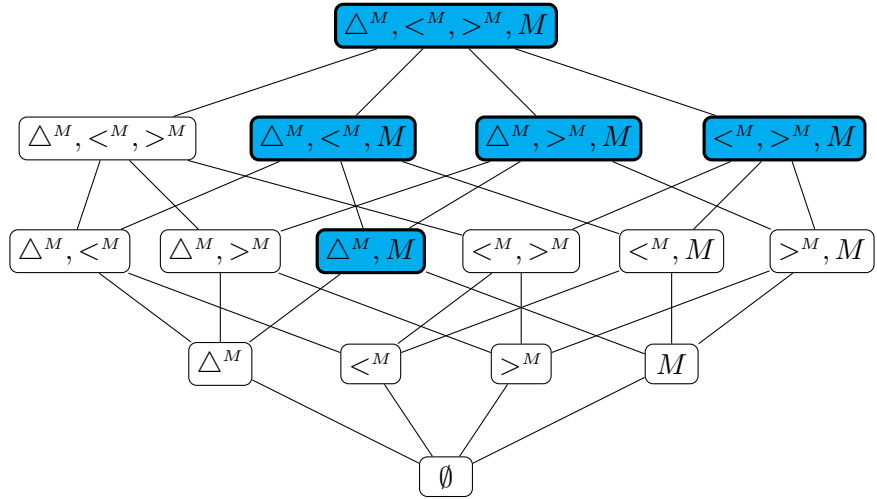

Figure 5: Complexity of finite-valued VCSPs $\mathcal{A}_{\mathbb{Q}_+}(S), S \subseteq \{\triangle^M, <^M, >^M, M\}$.

The complexity of combinations of costs from $\{\triangle^M, <^M, >^M, M\}$ are depicted in Figure 5: white nodes represent tractable cases and shaded nodes represent intractable cases.

The most interesting case is $\mathcal{A}_{\mathbb{Q}_+}(\{>^M, M\})$, which turns out to be tractable by a reduction to maximum *weighted* matching and hence is a proper generalization of class $\mathcal{A}_{\{0,1\}}(\{>, 1\})$.

**Proposition 3.34.** $\mathcal{A}_{\mathbb{Q}_+}(\{>^M, M\})$ *is tractable.*

*Proof.* The proof is similar to the proof of Proposition 3.20. Consider an instance $I$ in $\mathcal{A}_{\mathbb{Q}_+}(\{>^M, M\})$, and let

$$\alpha_{ij} = \min\{c_i(u) + c_{ij}(u, v) + c_j(v) \mid u \in D_i, v \in D_j\}$$

with the minimum being attained when $u = a_i^j$ and $v = a_j^i$. We can assume, without loss of generality, that the unary cost functions satisfy $\forall i$, $\exists d_i \in D_i$ such that $c_i(d_i) = 0$ (by subtracting, if necessary, $\min c_i(u)$ from the unary cost function $c_i$). This implies that $\alpha_{ij} \leq c_{ij}(d_i, d_j) \leq M$.

Suppose that $(b_i, b_j) \neq (a_i^j, a_j^i)$ and $c_{ij}(b_i, b_j) < M$. Then we can replace $(b_i, b_j)$ by $(a_i^j, a_j^i)$ in any solution to produce a solution of no greater cost: this is because all other binary costs involving $b_i$ or $b_j$ are necessarily maximal (i.e. equal to $M$). Therefore, setting $c_{ij}(b_i, b_j) = M$ does not change the cost of an optimal solution to the instance $I$. It follows that we can assume that there is at most one non-maximal binary cost $c_{ij}(a_i^j, a_j^i)$ in each binary cost function $c_{ij}$.

Consider the weighted complete graph $G$ with vertices $1, \ldots, n$ and edge weights $M - \alpha_{ij}$. Let $M_G$ be a maximum weighted matching of $G$. Define a solution $\mathbf{x} = \langle x_1, \ldots, x_n \rangle$ to $I$ by

$$x_i = \begin{cases} a_i^j & \text{if } \{i, j\} \in M_G \\ d_i & \text{otherwise} \end{cases}.$$

This solution is well-defined since $M_G$ is a matching. The weight of $M_G$ is

$$\sum_{\{i,j\} \in M_G} (M - \alpha_{ij}) \ = \ \binom{n}{2} M - \text{cost}(\mathbf{x}).$$

On the other hand, consider any solution $\mathbf{y}$ to $I$. Let

$$E(\mathbf{y}) \ = \ \{\{i, j\} \mid y_i = a_i^j \wedge y_j = a_j^i \wedge \alpha_{ij} < M\}.$$

$E(\mathbf{y})$ is a matching of $G$ of weight

$$\sum_{\{i,j\} \in E(\mathbf{y})} (M - \alpha_{ij}) \ \geq \binom{n}{2} M - \text{cost}(\mathbf{y}).$$

Since $M_G$ is a maximum weighted matching, we can deduce that $\text{cost}(\mathbf{y}) \geq \text{cost}(\mathbf{x})$. Hence $\mathbf{x}$ is an optimal solution.

Tractability follows from the tractability of the maximum weighted matching problem (Edmonds, 1965a). □

**Remark 3.35.** We have seen in the proof of Proposition 3.34 that $\mathcal{A}_{\mathbb{Q}_+}(\{>^M, M\})$ is tractable via a reduction to the maximum weighted matching problem (Edmonds, 1965a).

Similarly to Remark 3.21, it is easy to show that, conversely, any instance of the maximum weighted matching problem can be modelled as a VCSP instance from $\mathcal{A}_{\mathbb{Q}_+}(\{>^M, M\})$.

**Corollary 3.36** (of Theorem 3.1). $\mathcal{A}_{\mathbb{Q}_+}(\{<^M, M\})$ *is tractable.*

**Proposition 3.37.** $\mathcal{A}_{\mathbb{Q}_+}(\{\triangle^M, <^M, >^M\})$ *is tractable.*

*Proof.* Analogously to the proof of Proposition 3.30, instances contain at most 5 variables. □

**Proposition 3.38.** $\mathcal{A}_{\mathbb{Q}_+}(\{<^M, >^M, M\})$ *is intractable unless $|D| \leq 1$.*

*Proof.* By reduction from MAX-CUT as in the proof of Proposition 3.18 $\qquad\square$

**Proposition 3.39.** $\mathcal{A}_{\mathbb{Q}_+}(\{\triangle^M, M\})$ *is intractable unless* $|D| \leq 1$.

*Proof.* We will show intractability by reduction from VCSP on bipartite graphs and with Boolean domains which is known to be NP-hard (Cooper & Živný, 2011b). It suffices to replace all zero costs by $M$ in the reduction from VCSP on bipartite graphs given in the proof of Proposition 3.32 to produce an equivalent instance in $\mathcal{A}_{\mathbb{Q}_+}(\{\triangle^M, M\})$. $\qquad\square$

**Theorem 3.40.** *For* $|D| \geq 2$, *a class of binary finite-valued VCSP instances defined as* $\mathcal{A}_{\mathbb{Q}_+}(S)$, *where* $S \subseteq \{\triangle^M, <^M, >^M, M\}$, *is tractable if and only if* $S \subseteq \{<^M, M\}$ *or* $S \subseteq \{>^M, M\}$ *or* $S \subseteq \{\triangle^M, <^M, >^M\}$.

## 3.5 General-Valued VCSP

In this section, we focus on general-valued VCSPs. In other words, we consider the complete valuation structure $\overline{\mathbb{Q}}_+$ as the set of possible costs $\Omega$. In fact, the complexity classifications coincide with the classifications for finite-valued VCSPs obtained in Section 3.4.

Theorem 3.27 applies to $\Omega = \overline{\mathbb{Q}}_+$ as well. Indeed, the hard cases remain intractable when we allow more triangles (involving infinite costs), and the only tractable case, $\mathcal{A}_{\mathbb{Q}_+}(\{<, =\})$, remains tractable: $\mathcal{A}_{\overline{\mathbb{Q}}_+}(\{<, =\})$ is tractable by Theorem 3.1.

**Theorem 3.41.** *For* $|D| \geq 2$, *a class of binary general-valued VCSP instances defined as* $\mathcal{A}_{\overline{\mathbb{Q}}_+}(S)$, *where* $S \subseteq \{\triangle, <, >, =\}$, *is tractable if and only if* $S \subseteq \{<, =\}$.

Similarly with Theorem 3.33. Indeed, intractable cases remain intractable, and tractable cases remain tractable.

**Theorem 3.42.** *For* $|D| \geq 2$, *a class of binary general-valued VCSP instances defined as* $\mathcal{A}_{\overline{\mathbb{Q}}_+}(S)$, *where* $S \subseteq \{\triangle^0, <^0, >^0, 0\}$, *is tractable if and only if* $S \subseteq \{<^0, 0\}$, $S \subseteq \{>^0, 0\}$ *or* $S \subseteq \{\triangle^0, <^0, >^0\}$.

Similarly with Theorem 3.40. Indeed, intractable cases remain intractable, and tractable cases remain tractable. (The class $\mathcal{A}_{\overline{\mathbb{Q}}_+}(\{>^M, M\})$ becomes trivially tractable if $M = \infty$ as there is no solution of finite cost in instances with more than two variables.)

**Theorem 3.43.** *For* $|D| \geq 2$, *a class of binary general-valued VCSP instances defined as* $\mathcal{A}_{\overline{\mathbb{Q}}_+}(S)$, *where* $S \subseteq \{\triangle^M, <^M, >^M, M\}$, *is tractable if and only if* $S \subseteq \{<^M, M\}$ *or* $S \subseteq \{>^M, M\}$ *or* $S \subseteq \{\triangle^M, <^M, >^M\}$.

# 4. Cross-Free and Convex VCSPs

In Section 3, we studied the computational complexity of several classes of binary VCSPs. In all considered cases, the joint-winner property (JWP) was either the only one or one of only a few tractable cases.

In this section, we will generalise JWP to the *cross-free convexity property* (CFC). This property defines a novel tractable class for which we describe an efficient algorithm. In Section 4.4, we show that the neither of the two conditions in the definition of the CFC property can be dropped without rendering the problem NP-hard. Moreover, in Section 4.5, we present an extension of the CFC over Boolean domains. Section 4.6 is devoted to a related idea of overlaps studied previously only for SAT and Max-SAT.

## 4.1 Definition and Examples of Cross-Free and Convex VCSPs

A function $g : \{0, \ldots, s\} \to \overline{\mathbb{Q}}_+$ is called *convex on the interval* $[l, u]$ if $g$ is finite-valued on the interval $[l, u]$ and the derivative of $g$ is non-decreasing on $[l, u]$, i.e. $g(m+2) - g(m+1) \geq g(m+1) - g(m)$ for all $m = l, \ldots, u - 2$. For brevity, we will often say that $g$ is *convex* if it is convex on some interval $[l, u] \subseteq [0, s]$ and infinite elsewhere (i.e. on $[0, l-1] \cup [u+1, s]$).

Two sets $A_1, A_2 \subseteq A$ are said to be *nested* if they are either disjoint or one is a subset of the other (i.e. $A_1 \cap A_2 = \emptyset$, $A_1 \subseteq A_2$ or $A_2 \subseteq A_1$). If $A_1$ and $A_2$ are not nested, then we say that they *overlap*. We say that $A_1$ and $A_2$ *incompletely overlap* if $A_1$ and $A_2$ overlap and $A_1 \cup A_2 \neq A$.

Sets $A_1, \ldots, A_r$ are called *laminar* (Schrijver, 2003) (or *hierarchically nested*; see Cooper & Živný, 2011a) if for any $1 \leq i, j \leq r$, $A_i$ and $A_j$ are nested. Sets $A_1, \ldots, A_r \subseteq A$ are called *cross-free* if for every $1 \leq i, j \leq r$, either $A_i \subseteq A_j$, or $A_i \supseteq A_j$, or $A_i \cap A_j = \emptyset$, or $A_i \cup A_j = A$ (Schrijver, 2003). It is clear that if sets $A_1, \ldots, A_r$ are laminar, then $A_1, \ldots, A_r$ are also cross-free.

For notational convenience, we interpret a solution $\mathbf{x}$ (i.e. an assignment to the variables $v_1, \ldots, v_n$) to a VCSP instance as the set of $\langle$variable,value$\rangle$ assignments $\{\langle v_i, x_i \rangle \mid x_i \in D_i \wedge i = 1, \ldots, n\}$.

If $A_i$ is a set of $\langle$variable,value$\rangle$ assignments of a VCSP instance $\mathcal{P}$ and $\mathbf{x}$ a solution to $\mathcal{P}$, then we use the notation $|\mathbf{x} \cap A_i|$ to represent the number of $\langle$variable,value$\rangle$ assignments in the solution $\mathbf{x}$ which lie in $A_i$.

**Definition 4.1** (Laminar/Cross-free convexity). *Let $\mathcal{P}$ be a VCSP instance. Let $A_1, \ldots, A_r$ be laminar (cross-free) sets of $\langle$variable,value$\rangle$ assignments of $\mathcal{P}$. Let $s_i$ be the number of distinct variables occurring in the set of $\langle$variable,value$\rangle$ assignments $A_i$. Instance $\mathcal{P}$ satisfies the laminar-free (cross-free) convexity property if the objective function of $\mathcal{P}$ is $g(\mathbf{x}) = g_1(|\mathbf{x} \cap A_1|) + \ldots + g_r(|\mathbf{x} \cap A_r|)$ where each $g_i : [0, s_i] \to \mathbb{Q}_+$ $(i = 1, \ldots, r)$ is convex on an interval $[l_i, u_i] \subseteq [0, s_i]$ and $g_i(z) = \infty$ for $z \in [0, l_i - 1] \cup [u_i + 1, s_i]$.*

We remark that the functions $g_i$ in Definition 4.1 are not the cost functions associated with the constraints.

It follows from the definition that the laminar convexity property implies the cross-free convexity property.

**Remark 4.2.** Observe that the addition of any unary cost function cannot destroy the laminar or cross-free convexity property. This is because for each $\langle$variable,value$\rangle$ assignment $\langle v_j, a \rangle$ we can add the singleton $A_i = \{\langle v_j, a \rangle\}$ which is necessarily either disjoint from or a subset of any other set $A_k$ (and furthermore the corresponding function $g_i : \{0, 1\} \to \mathbb{Q}_+$ is trivially convex).

We now give a very special case of the cross-free convexity property, where all sets are disjoint and thus trivially cross-free.

**Example 4.3** (Value-based soft GCC). The GLOBAL CARDINALITY CONSTRAINT (GCC), introduced by Régin (1996), is a generalisation of the ALLDIFFERENT constraint (Régin, 1994). Given a set of $n$ variables, the GCC specifies for each domain value $d$ a lower bound $l_d$ and an upper bound $u_d$ on the number of variables that are assigned value $d$. The ALLDIFFERENT constraint is the special case of GCC with $l_d = 0$ and $u_d = 1$ for every $d$. Soft versions of the GCC have been considered by van Hoeve, Pesant, & Rousseau (2006).

The value-based soft GCC minimises the number of values below or above the given bound. We show that the value-based soft GCC satisfies the cross-free convexity property.

For every domain value $d \in D$, let $A_d = \{\langle v_i, d \rangle : i = 1, \ldots, n\}$. Clearly, $A_1, \ldots, A_s$ are disjoint, where $s = |D|$. For every $d$, let

$$g_d(m) = \begin{cases} l_d - m & \text{if } m < l_d \\ 0 & \text{if } l_d \le m \le u_d \\ m - u_d & \text{if } m > u_d \end{cases}$$

It follows readily from the definition of $g_d$ that the sequence $g_d(m+1) - g_d(m)$, for $m = 0, \ldots, n-1$, is the sequence $-1, \ldots, -1, 0, \ldots, 0, 1, \ldots, 1$. Therefore, for every $d$, $g_d$ has a non-decreasing derivative and hence is convex.

**Example 4.4** (Nested value-based soft GCC). Being able to nest GCC constraints is useful in many staff assignment problems where there is a hierarchy (e.g. senior manager-manager-personnel, foreman-worker, or senior nurse-nurse) (Zanarini & Pesant, 2007). We might want to impose soft global cardinality constraints such as each day we prefer that there are between 10 and 15 people at work, of which at least 5 are managers among whom there is exactly 1 senior manger, with convex penalties as described in Example 4.3 if these constraints do not hold.

Suppose that the constraints of a VCSP instance consist of soft GCC constraints on pairwise nested sets of variables $S_1, \ldots, S_t$. Let $A_{id} = \{\langle x, d \rangle : x \in S_i\}$. Clearly, the sets of assignments $A_{id}$ are cross-free and, as shown in Example 4.3, the cost functions corresponding to each soft GCC constraint are convex.

The main result of this section is the following theorem:

**Theorem 4.5.** *Any VCSP instance $\mathcal{P}$ satisfying the cross-free convexity property can be solved in polynomial time.*

Firstly, we present an algorithm to solve VCSPs satisfying the laminar convexity property, followed by a reduction from the cross-free case to the laminar case. Secondly, we give a proof of polynomial-time complexity of this algorithm.

## 4.2 Algorithm for Laminar Convex VCSPs

We call the sets $A_i$ ($i = 1, \ldots, r$) assignment-sets. We assume that the assignment-sets $A_i$ are distinct, since if $A_i = A_j$ then these two sets can be merged by replacing the two functions $g_i, g_j$ by their sum (which is necessarily also convex). Without loss of generality, we can assume that the assignment-set consisting of all variable-value assignments is present, and the corresponding function is the constant zero function. (If the corresponding function $g_i$ is not the constant zero function, then we just add the constant term $g_i(n)$ to the objective function.) This will be useful in the construction described below. We say that assignment-set $A_k$ is the *father* of assignment-set $A_i$ if it is the minimal assignment-set which properly contains $A_i$, i.e. $A_i \subset A_k$ and $\nexists A_j$ such that $A_i \subset A_j \subset A_k$. It follows from the definition of laminarity that $A_k$ is unique and hence that the father relation defines a tree. Moreover, again from the definition of laminarity, for every variable $v_i$ of $\mathcal{P}$ and every $a \in D_i$, there is a unique minimal assignment-set containing $\langle v_i, a \rangle$.

We construct a directed graph $G_{\mathcal{P}}$ whose minimum-cost integral flows of value $n$ are in one-to-one correspondence with the solutions to $\mathcal{P}$. $G_{\mathcal{P}}$ has the following nodes:

1. the source node $s$;

2. a variable node $v_i$ $(i = 1, \ldots, n)$ for each variable of $\mathcal{P}$;

3. an assignment node $\langle v_i, d \rangle$ $(d \in D_i, \ i = 1, \ldots, n)$ for each possible variable-value assignment in $\mathcal{P}$;

4. an assignment-set node $A_i$ $(i = 1, \ldots, r)$ for each assignment-set in $\mathcal{P}$;

5. the sink node $t$, which we identify with the assignment-set consisting of all variable-value assignments.

$G_{\mathcal{P}}$ has the following arcs:

1. $a = (s, v_i)$ for each variable $v_i$ of $\mathcal{P}$; the demand and capacity are given by $d(a) = c(a) = 1$ (this forces a flow of exactly 1 through each variable node $v_i$); the weight function is given by $w(a) = 0$;

2. $a = (v_i, \langle v_i, d \rangle)$ for all variables $v_i$ and for each $d \in D_i$; $d(a) = 0$; $c(a) = 1$; $w(a) = 0$;

3. $a = (\langle v_i, d \rangle, A_j)$ for all variables $v_i$ and for each $d \in D_i$, where $A_j$ is the minimal assignment-set containing $\langle v_i, d \rangle$; $d(a) = 0$; $c(a) = 1$; $w(a) = 0$;

4. for each assignment-set $A_i$ with father $A_j$, there is an arc $a$ from $A_i$ to $A_j$ with weight function $g_i$, demand $d(a) = l_i$ and capacity $c(a) = u_i$.

Clearly, $G_{\mathcal{P}}$ can be constructed from $\mathcal{P}$ in polynomial time. We now prove that minimum-cost flows $f$ of value $n$ in $G_{\mathcal{P}}$ are in one-to-one correspondence with solutions to $\mathcal{P}$ and, furthermore, that the cost of $f$ is equal to the cost in $\mathcal{P}$ of the corresponding solution.

All feasible flows have value $n$ since all $n$ arcs $(s, v_i)$ leaving the source have both demand and capacity equal to 1. Flows in $G_{\mathcal{P}}$ necessarily correspond to the assignment of a unique value $x_i$ to each variable $v_i$ since the flow of 1 through node $v_i$ must traverse a node $\langle v_i, x_i \rangle$ for some unique $x_i \in D_i$. It remains to show that for every assignment $\mathbf{x} = \{\langle v_1, x_1 \rangle, \ldots, \langle v_n, x_n \rangle\}$ which is feasible (i.e. whose cost in $\mathcal{P}$ is finite), there is a corresponding minimum-cost feasible flow $f$ in $G_{\mathcal{P}}$ of cost $g(x) = g_1(|\mathbf{x} \cap A_1|) + \ldots + g_r(|\mathbf{x} \cap A_r|)$.

For each arc $a$ that is incoming to or outgoing from $\langle v_i, d \rangle$ in $G_{\mathcal{P}}$, let $f(a) = 1$ if $d = x_i$ and 0 otherwise. By construction, each assignment-set node $A_i$ in $G_{\mathcal{P}}$ has exactly one outgoing arc to its father assignment-set. The flow $f_a$ in arc $a$ from $A_i$ to its father assignment-set $A_j$ is uniquely determined by the assignment of values to variables in the solution $\mathbf{x}$. Trivially, this is therefore a minimum-cost flow corresponding to the assignment $\mathbf{x}$. The cost of flow $f$ is clearly $\sum_i g_i(|\mathbf{x} \cap A_i|)$ which corresponds precisely to the cost of the assignment $\mathbf{x}$.

Having proved the correspondence between the cost of solutions to $\mathcal{P}$ and the cost of minimum-cost flows, it follows that the algorithm, which for given $\mathcal{P}$ constructs $G_{\mathcal{P}}$ and finds a minimum-cost flow, is correct.

**Example 4.6.** Let $\mathcal{P}$ be a VCSP instance with 4 variables $v_1, v_2, v_3, v_4$, $D_1 = D_2 = D_3 = D_4 = \{0, 1\}$, and the assignment-sets $A_i$, $1 \leq i \leq 8$ given in Figure 6. The cost functions $g_i$, $1 \leq i \leq 8$ are arbitrary convex functions.

The network $G_{\mathcal{P}}$ corresponding to instance $\mathcal{P}$ is shown in Figure 7: demands and capacities are in square brackets for the corresponding layer of the graph, and weights of arcs without numbers are 0. The only non-zero weight functions are on arcs between assignment-sets; those arcs have the corresponding cost functions $g_i$, $1 \leq i \leq 7$. Set $A_8$ is identified with the sink $t$. Minimum-cost feasible flows in $G_{\mathcal{P}}$ correspond to assignments to $\mathcal{P}$ modulo the addition of the constant $g_8(4)$ (since there are 4 variables and $A_8$ consists of all variable-value assignments). The bold red edges represent flow $f$ corresponding to the assignment $v_1 = v_2 = 1$ and $v_3 = v_4 = 0$ with the total cost $g_1(1) + g_2(0) + g_3(2) + g_4(1) + g_5(0) + g_6(1) + g_7(3)$. Finding a minimum-cost flow in $G_{\mathcal{P}}$ is equivalent to finding an optimal solution to $\mathcal{P}$.

## 4.3 From Laminar VCSPs to Cross-Free VCSPs

An alternative way of expressing the definition of cross-freeness is that for every $1 \leq i, j \leq r$, one of $A_i \cap (A \setminus A_j)$, $(A \setminus A_i) \cap A_j$, $A_i \cap A_j$, $(A \setminus A_i) \cap (A \setminus A_j)$ is empty. It follows directly that if $A_1, \ldots, A_r$ are cross-free then so are $A_1, \ldots, A_r$, $(A \setminus A_i)$ for any $1 \leq i \leq r$.

We now show how to reduce any VCSP instance with the cross-free convexity property to an instance satisfying the laminar convexity property.

First we show that without loss of generality, we can assume that every $A_i$ satisfies $|A_i| \leq \lfloor |A|/2 \rfloor$, $1 \leq i \leq r$. Let $A_i$ be arbitrary such that $|A_i| > \lfloor |A|/2 \rfloor$. As pointed out above, without loss of generality there is $A_j$, $1 \leq j \leq r$, such that $A_j = A \setminus A_i$. (If there is no such $A_j$ among $A_1, \ldots, A_r$, we can add $A_j$ with the corresponding convex cost function being the constant zero cost function. This would only double the number of assignment-sets.)

Let $h_i$ be defined by $h_i(y) = g_i(n - y)$ and let $g'_j = g_j + h_i$. Clearly $g'_j$ is convex, and furthermore

$$
\begin{aligned}
g'_j(|A_j \cap x|) &= g_j(|A_j \cap x|) + h_i(|A_j \cap x|) \\
&= g_j(|A_j \cap x|) + g_i(n - |A_j \cap x|) \\
&= g_j(|A_j \cap x|) + g_i(|A \cap x| - |A_j \cap x|) \\
&= g_j(|A_j \cap x|) + g_i(|A_i \cap x|).
\end{aligned}
$$

So we can eliminate the set $A_i$ and its cost function $g_i$ by replacing $g_i, g_j$ by a single cost function $g'_j$.

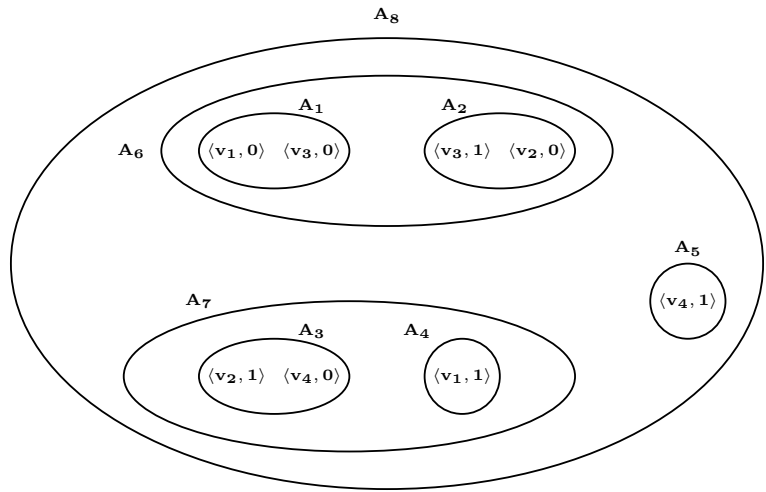

Figure 6: Laminar sets of assignments from instance $\mathcal{P}$ of Example 4.6.

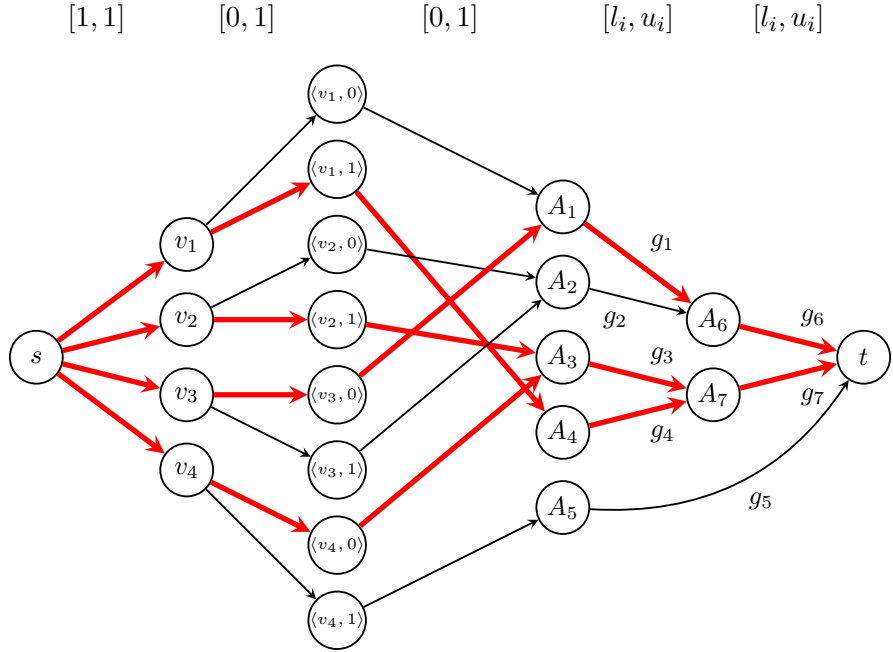

Figure 7: Network $G_{\mathcal{P}}$ corresponding to the VCSP $\mathcal{P}$ of Example 4.6.

Since all sets are at most half of the size of $A$, if $A_i \cup A_j = A$ for some $1 \le i, j \le r$, then necessarily $A_j = A \setminus A_i$. However, in this case, using the same argument as above, for each such pair of complementary sets $A_i$ and $A_j$, we can eliminate $A_i$ and its cost function $g_i$ by replacing $g_i, g_j$ by a single cost function $g'_j$. Consequently, the resulting sets $A_1, \ldots, A_r$ are laminar.

**Complexity**   Let $\mathcal{P}$ be a VCSP instance with $n$ variables, each with a domain of size at most $d$, and $r$ laminar assignment-sets $A_i$. The maximum number of distinct non-overlapping sets $A_i$ is $2nd - 1$ since the sets of assignments $A_i$ form a tree with at most $nd$ leaves (corresponding to single ⟨variable,value⟩ assignments) and in which all non-leaf nodes have at least two sons. Thus $r = O(nd)$. The network $G_{\mathcal{P}}$ has $n' = O(n + nd + r) = O(nd)$ vertices and arcs. $G_{\mathcal{P}}$ can be built in $O((nd)^2)$ time in a top-down manner, by adding assignment-sets in inverse order of size (which ensures that an assignment-set is always inserted after its father) and using a table $T[⟨v, a⟩] =$smallest assignment-set (in the tree being built) containing ⟨v, a⟩.

In a network with $n'$ vertices and $m'$ arcs with capacities at most $U$, the minimum convex cost flow problem can be solved in time $O((m' \log U) SP(n', m'))$, where $SP(n', m')$ is the time to compute a shortest directed path in a network with $n'$ vertices and $m'$ edges (Ahuja, Magnanti, & Orlin, 2005). Using Fibonacci heaps (Fredman & Tarjan, 1987), $SP(n', m') = O(m' + n' \log n') = O(nd \log(nd))$, since the number of vertices $n'$ and arcs $m'$ are both $O(nd)$. The maximum capacity $U$ in the network $G_{\mathcal{P}}$ is at most $n$. Hence an optimal solution to a cross-free convex VCSP can be determined in $O((nd \log n)(nd \log(nd))) = O((nd)^2(\log n)(\log n + \log d))$ time.

**Remark 4.7.** In our previous work (Cooper & Živný, 2011b), we proved a special case of Theorem 4.5 where all functions $g_i$, $1 \le i \le r$, are non-decreasing and assignment sets are laminar. (Previously, the laminar convexity property for non-decreasing functions $g_i$, $1 \le i \le r$, was called the non-overlapping convexity property; also, assignment-sets were called assignment-cliques; see Cooper & Živný, 2011b.)

The presented algorithm is similar to the algorithm of Cooper & Živný (2011b) based on finding a minimum-cost flow in a network. The main difference is that we require only a single arc between any pair of nodes and the corresponding cost function $g_i$ is now an arbitrary convex function (which is not necessarily non-decreasing). The running time of our algorithm is thus better than the running time of the algorithm from our previous work (Cooper & Živný, 2011b), which is $O(n^3 d^2)$. The improvement is mostly due to the fact that the new construction involves only $O(nd)$ arcs as opposed to $O((nd)^2)$ arcs in the previous work (Cooper & Živný, 2011b). Moreover, our algorithm solves a strictly bigger class of problems compared to the previous result (Cooper & Živný, 2011b). Overall, we solve more and faster!

**Remark 4.8.** We remark that since our construction is projection-safe (Lee & Leung, 2009), it can be used for Soft Global Arc Consistency for cross-free convex constraints.

**Remark 4.9.** For a VCSP instance $\mathcal{P}$ with the objective function of the form $g(\mathbf{x}) = \sum_{i=1}^r g_i(|\mathbf{x} \cap A_i|)$, it follows from the definitions that we can test in polynomial time whether or not $\mathcal{P}$ satisfies the cross-free convexity property; that is, whether $g_i$ are convex and $A_i$ are cross-free, for $1 \leq i \leq r$. In fact, the described algorithm requires that the assignment-sets $A_i$ and the functions $g_i$ be given explicitly.

In the conference version of this work (Cooper & Živný, 2011a), we mentioned the recognition problem as an open problem. In fact, this problem is easily shown intractable. Given an arbitrary VCSP instance $\mathcal{P}$, there always exists a cross-free convex instance $\mathcal{P}'$ whose optimal solution coincides with a fixed optimal solution to $\mathcal{P}$. Therefore, finding $\mathcal{P}'$ is impossible in polynomial time unless P=NP as otherwise an arbitrary VCSP instance $\mathcal{P}$ could be solved in polynomial time using Theorem 4.5.

### 4.4 Maximality of Cross-Free Convexity

This section shows that relaxing either convexity or cross-freeness (in fact, laminarity) in Definition 4.1 leads to intractability.

**Theorem 4.10.** *The class of VCSP instances whose objective function is of the form $g(\mathbf{x}) = g_1(|\mathbf{x} \cap A_1|) + \ldots + g_r(|\mathbf{x} \cap A_r|)$ where the functions $g_i$ are convex, but the sets of assignments $A_i$ may overlap, is NP-hard, even if $|A_i| \leq 2$ for all $i \in \{1, \ldots, r\}$ and all variables are Boolean.*

*Proof.* It suffices to demonstrate a polynomial-time reduction from the well-known NP-hard problem MAX-2SAT (Garey & Johnson, 1979). Any MAX-2SAT clause $l_1 \vee l_2$ (where $l_1, l_2$ are literals) is equivalent to the convex cost function $g(|\mathbf{x} \cap \{l_1, l_2\}|)$ where $g(0) = 1$ and $g(1) = g(2) = 0$. It is therefore possible to code any instance of MAX-2SAT using convex cost functions (on possibly overlapping sets of assignments). $\square$

**Theorem 4.11.** *The class of VCSP instances whose objective function is of the form $g(\mathbf{x}) = g_1(|\mathbf{x} \cap A_1|) + \ldots + g_r(|\mathbf{x} \cap A_r|)$ where the sets of assignments $A_i$ are laminar, but the functions $g_i$ are not necessarily convex, is NP-hard even if $|A_i| \leq 3$ for all $i \in \{1, \ldots, r\}$ and all variables are Boolean.*

*Proof.* We give a polynomial-time reduction from the well-known NP-complete problem 3SAT (Garey & Johnson, 1979). Let $I_{3SAT}$ be an instance of 3SAT with $m$ clauses. The constraint ALLEQUAL$(l_1, l_2, l_3)$ (where $l_1, l_2, l_3$ are literals) is equivalent to the (non-convex) cost function $g(|\mathbf{x} \cap \{l_1, l_2, l_3\}|)$ where $g(0) = g(3) = 0$ and $g(1) = g(2) = \infty$. For each variable $v$ in $I_{3SAT}$, we use the following gadget $G_v$ based on non-overlapping ALLEQUAL constraints to produce multiple copies $v_1, \ldots, v_m$ of the variable $v$ and multiple copies $w_1, \ldots, w_m$ of its negation $\overline{v}$: $G_v$ consists of the constraints ALLEQUAL$(\overline{u_i}, \overline{v_i}, \overline{y_i})$ ($i \in \{1, \ldots, m\}$), ALLEQUAL$(y_i, \overline{w_i}, u_{i+1})$ ($i \in \{1, \ldots, m-1\}$), and ALLEQUAL$(y_m, \overline{w_m}, u_1)$, where the variables $u_i, y_i$ occur only in the gadget $G_v$. It is easy to verify that $G_v$ imposes $v_1 = \ldots = v_m = \overline{w_1} = \ldots = \overline{w_m}$. Furthermore, the variables $v_i, w_i$ occur only negatively in $G_v$. We now replace the $i$th clause of $I_{3SAT}$ by a clause in which each positive variable $v$ is replaced by its $i$th copy $v_i$ and each negative variable $\overline{v}$ is replaced by the $i$th copy $w_i$ of $\overline{v}$. This produces a laminar VCSP instance which is equivalent to $I_{3SAT}$ (but whose cost functions are not all convex). $\square$

Note that the NP-hardness reduction in the proof of Theorem 4.11 requires assignment-sets of size up to 3. This leaves open the complexity of laminar (and cross-free) non-convex VCSPs where all assignment-sets are of size at most 2.

The following result shows that the complexity of cross-free non-convex VCSPs with assignment-sets of size 2 and domains of size $d$ is polynomial-time equivalent to cross-free non-convex VCSPs with assignment-sets of size 2 and domains of size at most 3.

**Proposition 4.12.** *Cross-free VCSPs with assignment-sets of size at most 2 and domains of size $d > 3$ are polynomial-time equivalent to cross-free VCSPs with assignment-sets of size at most 2 and domains of size at most 3.*

*Proof.* First we observe that for VCSPs with assignment-sets of size at most 2, laminarity and cross-freeness are almost identical. The extra condition in the definition of cross-freeness (for $A_1, A_2 \subseteq A$, $A_1 \cap A_2 \neq \emptyset \Rightarrow A_1 \cup A_2 = A$) is irrelevant for instances with more than 3 variable-value assignments. Hence we only need to prove the equivalence for laminar VCSPs.

Let $v_\ell$ be such that $D_\ell = \{a_1, \ldots, a_k\}$, where $k > 3$. We replace $v_\ell$ by $k$ variables $v_{\ell,1}, \ldots, v_{\ell,k}$ with respective domains $D_{\ell,1} = \{1, a_1\}$, $D_{\ell,i} = \{0, 1, a_i\}$ for $i = 2, \ldots, k-1$, and $D_{\ell,k} = \{0, a_k\}$. (Here we assume, without loss of generality, that 0 and 1 are different from $a_i$, $i = 1, \ldots, k$.) Moreover, we introduce $k-1$ new assignment-sets $\{\langle v_{\ell,i}, 1 \rangle, \langle v_{\ell,i+1}, 0 \rangle\}$ for $i = 1, \ldots, k-1$ with the associated convex function $g$ defined as $g(1) = 0$ and $g(0) = g(2) = \infty$. Finally, in any assignment-set involving the variable-value assignment $\langle v_\ell, a_i \rangle$ (for some $i \in \{1, \ldots, k\}$), this assignment is replaced by $\langle v_{\ell,i}, a_i \rangle$.

The function $g$ applied to the assignment-sets $\{\langle v_{\ell,i}, 1 \rangle, \langle v_{\ell,i+1}, 0 \rangle\}$ ensures that the only possible finite-cost assignments to variables $v_{\ell,1}, \ldots, v_{\ell,k}$ are of the form $1, \ldots, 1, a_i, 0, \ldots, 0$. Since exactly one of the variables $v_{\ell,1}, \ldots, v_{\ell,k}$ is assigned a value from $D_\ell$, there is a one-to-one correspondence between optimal solutions to the transformed instance and the original instance. $\square$

The tractability of cross-free non-convex VCSPs with assignment-sets of size 2 over domains of size 3 (or larger, by Proposition 4.12) is left as an open problem.

The case of cross-free assignment-sets of size at most 2 over Boolean domains is shown tractable in Theorem 4.21 in Section 4.6.

## 4.5 Renamable Boolean Cross-Free Convex VCSPs

In this section we extend the class of cross-free convex VCSPs to allow renaming of certain variables in the case of Boolean domains. In this section we will consider only Boolean VCSPs.

We begin by illustrating the notion of renaming by means of an example. First, we require some notation. Cost function $\text{ATMOST}_r(A)$ returns 0 if $\mathbf{x}$ contains at most $r$ assignments from the set of assignments $A$, and $\text{ATMOST}_r(A)$ returns 1 otherwise. Similarly, cost function $\text{ATLEAST}_r(A)$ returns 0 if $\mathbf{x}$ contains at least $r$ assignments from the set of assignments $A$, and $\text{ATLEAST}_r(A)$ returns 1 otherwise. Note that cost functions $\text{ATLEAST}_1$ and $\text{ATMOST}_r$, where $r = |A| - 1$, are both convex on $[0, |A|]$.

**Example 4.13.** Let $\mathcal{P}$ be a MAX-SAT instance given in CNF form by the following clauses:

$$(a \vee b \vee c), \quad (c \vee d), \quad (\neg c \vee \neg d \vee e), \quad (\neg a \vee \neg e).$$

Clearly, a clause with literals $A$ can be written as $\textsc{AtLeast}_1(A)$ in the VCSP encoding of this instance. Notice that, in this example, the first two clauses are overlapping. However, we can replace the second clause by the equivalent constraint $\textsc{AtMost}_1(\{\neg c, \neg d\})$. This gives us an equivalent problem with the following constraints:

$$(a \vee b \vee c), \quad \textsc{AtMost}_1(\{\neg c, \neg d\}), \quad (\neg c \vee \neg d \vee e), \quad (\neg a \vee \neg e).$$

Now $\mathcal{P}$ is expressed as an instance satisfying the cross-free convexity property on the cross-free sets of assignments $\{a, b, c\}$, $\{\neg c, \neg d\}$, $\{\neg c, \neg d, e\}$, $\{\neg a, \neg e\}$.

Example 4.13 leads to the following definitions:

**Definition 4.14.** *Given a valued constraint in the form of the cost function $g(|\mathbf{x} \cap A|)$, where $A$ is a set of Boolean assignments (i.e. literals) of size $m$, we define the* renaming *of this valued constraint, on the set of Boolean assignments $\bar{A} = \{\neg \ell \mid \ell \in A\}$, as the valued constraint $g'(|\mathbf{x} \cap \bar{A}|) = g(m - |\mathbf{x} \cap \bar{A}|) = g(|\mathbf{x} \cap A|)$.*

The function $g'(z) = g(m - z)$ is clearly convex if and only if $g$ is convex.

**Definition 4.15.** *A Boolean VCSP instance $\mathcal{P}$ with the objective function $g_1(|\mathbf{x} \cap A_1|) + \ldots + g_r(|\mathbf{x} \cap A_r|)$ is* renamable cross-free convex *if there is a subset of the constraints of $\mathcal{P}$ whose renaming results in an equivalent VCSP instance $\mathcal{P}'$ which is cross-free convex.*

**Theorem 4.16.** *The class of renamable cross-free convex VCSPs is recognisable and solvable in polynomial time.*

*Proof.* We show that recognition is polynomial-time by a simple reduction to 2-SAT, a well-known problem solvable in polynomial time (Garey & Johnson, 1979). Let $\mathcal{P}$ be a Boolean VCSP instance with $r$ constraints such that the $i$th constraint ($i = 1, \ldots, r$) is $g_i(|\mathbf{x} \cap A_i|)$ for a convex function $g_i$. For each constraint in $\mathcal{P}$, there is a Boolean variable $ren_i$ indicating whether or not the $i$th constraint is renamed. For each pair of distinct $i, j \in \{1, \ldots, r\}$, we add clauses of length 2 as follows:

1. if $A_i$ and $A_j$ incompletely overlap then add constraint $ren_i \Leftrightarrow \neg ren_j$ (since we must rename just one of the two constraints);

2. if $\bar{A}_i$ and $A_j$ incompletely overlap then add constraint $ren_i \Leftrightarrow ren_j$ (to avoid introducing an overlap by a renaming).

It is easy to see that solutions to the constructed 2-SAT instance correspond to valid renamings of $\mathcal{P}$ which give rise to an equivalent VCSP instance satisfying the cross-free convexity property. Tractability of solving the resulting renamed instance follows directly from Theorem 4.5. $\square$

## 4.6 Knuth-Nested VCSPs

In order to relate our work to previous work, in this section we present a different class of tractable VCSPs which considers sets of variables (rather than sets of assignments) and allows overlaps of size 1. We show that a known tractable class can be extended from $\textsc{Max-SAT}$ to VCSPs. We then apply this result to show that in a very special case the assumption of convexity in cross-free convex VCSPs can be dropped.

**Definition 4.17.** *Given a VCSP instance $\mathcal{P}$ with variables $V = \{v_1, \ldots, v_n\}$ and constraints with scopes $C = \{C_1, \ldots, C_m\}$, we define the incidence graph of $\mathcal{P}$ as $I_{\mathcal{P}} = (V(I_{\mathcal{P}}), E(I_{\mathcal{P}}))$, where $V(I_{\mathcal{P}}) = V \cup C$ and $E(I_{\mathcal{P}}) = \{\{v_i, C_j\} \mid v_i \in C_j\}$.*

**Definition 4.18.** *A VCSP instance $\mathcal{P}$ is called* Knuth-nested *if the variables of $\mathcal{P}$ can be linearly ordered $v_1, \ldots, v_n$ such that $I_{\mathcal{P}}$ together with the edges $\{\{v_i, v_{i+1}\} \mid 1 \leq i \leq n\} \cup \{v_n, v_1\}$ allows a planar drawing so that the circle $v_1, \ldots, v_n, v_1$ bounds the outer face. $\mathcal{P}$ is called* Knuth-co-nested *if the constraint scopes of $\mathcal{P}$ can be linearly ordered $C_1, \ldots, C_m$ such that $I_{\mathcal{P}}$ together with the edges $\{\{C_i, C_{i+1}\} \mid 1 \leq i \leq m\} \cup \{C_m, C_1\}$ allows a planar drawing so that the circle $C_1, \ldots, C_m, C_1$ bounds the outer face.*

Knuth described a linear-time algorithm for solving Knuth-nested SAT instances (Knuth, 1990). Kratochvíl and Křivánek generalised Knuth's result and provided a linear-time algorithm for recognising and solving Knuth-nested and Knuth-co-nested SAT/MAX-SAT instances (Kratochvíl & Křivánek, 1993). Henderson in his Master's thesis showed several different proofs of these results, including a proof that Knuth-nested and Knuth-co-nested SAT/MAX-SAT instances have treewidth at most three (Biedl & Henderson, 2004; Henderson, 2005), and hence are solvable in polynomial time via a standard dynamic programming approach.

**Theorem 4.19.** *The class of Knuth-nested and Knuth-co-nested VCSP instances with constraints of bounded arity is recognisable and solvable in polynomial time.*

*Proof.* Recognition can be reduced, via a simple reduction from the work of Kratochvíl and Křivánek (1993), to the planarity testing problem (Hopcroft & Tarjan, 1974).

Following Henderson's argument (2005, p. 21), it is easy to show that if $\mathcal{P}$ is Knuth-nested or Knuth-co-nested, then the incidence graph $I_{\mathcal{P}}$ of $\mathcal{P}$ has treewidth at most 3. A VCSP with domains of size at most $d$, constraints of arity at most $k$ and incidence graph $I_{\mathcal{P}}$ is clearly equivalent to a binary VCSP with constraint graph $I_{\mathcal{P}}$ in which domains are of size at most $d^k$. The result then follows from the fact that any VCSP instance with a constraint graph of bounded treewidth is solvable in polynomial time (Bertelé & Brioshi, 1972). □

Note that the class of Knuth-nested (Knuth-co-nested) VCSP instances (in fact, even SAT instances) cannot be generalised as it follows from the work of Lichtenstein (1982) that the satisfiability of the conjunction of two Knuth-nested formulas is NP-complete.

We now show that the class of Knuth-nested/Knuth-co-nested instances from this section is incomparable with the class of cross-free convex instances defined in Section 4.1 even in the special case of Boolean formulas. Moreover, we also show that the class of Knuth-nested/Knuth-co-nested instances is incomparable with the class of renamable Boolean cross-free convex instances defined in Section 4.5.

**Example 4.20.** The SAT instance $I = (x \vee y) \wedge (y \vee z) \wedge (\neg y \vee w)$ is Knuth-nested and Knuth-co-nested, but neither cross-free nor renamable cross-free.

The following SAT instance is neither Knuth-nested nor Knuth-co-nested, but is cross-free (in fact laminar): $(x \vee y \vee z) \wedge (\neg x \vee u \vee v) \wedge (\neg y \vee \neg u \vee w) \wedge (\neg z \vee \neg v \vee \neg w)$.

We now turn our attention to cross-free VCSPs with possibly non-convex cost functions and assignment-sets of size at most 2. We show that the tractability of Boolean VCSP instances, i.e. instances over 2-element domains, follows from Theorem 4.19. Recall that the case of convex cost functions is tractable from Theorem 4.5, but that, from Theorem 4.11, the case of non-convex cost functions is intractable for assignments sets of size 3.

**Theorem 4.21.** *Any cross-free Boolean VCSP instance with assignment-sets of size at most 2 is solvable in polynomial time.*

*Proof.* As mentioned in the proof of Proposition 4.12, first observe that for VCSPs with assignment-sets of size 2 (or any fixed size for that matter) laminarity and cross-freeness are almost identical. The extra condition in the definition of cross-freenes (for $A_1, A_2 \subseteq A$, $A_1 \cap A_2 \neq \emptyset \Rightarrow A_1 \cup A_2 = A$) is irrelevant for instances with more than 3 variable-value assignments. Hence we only need to prove tractability for laminar Boolean VCSPs with assignment-sets of size at most 2.

We show that any Boolean VCSP with laminar assignment-sets of size at most 2 is Knuth-nested. Tractability then follows from Theorem 4.19.

Take an arbitrary variable, for instance $v_1$. We will show that there is an order $<$ of variables satisfying the requirements of the Knuth-nested property. Since $|D_1| = 2$, there are at most two assignment-sets, say $A_i$ and $A_j$, containing (different) assignments to $v_1$. Now since all assignment-sets are of size at most two, there is at most one more assignment in $A_i$, say an assignment to variable $v_k$. We define $v_1 < v_k$. Similarly, there is at most one more assignment in $A_j$, say an assignment to variable $v_l$, and we define $v_l < v_1$. Continuing the same reasoning for variables $v_l$ and $v_k$, we can get another variable smaller (in the order $<$ we are building) than $v_l$ and another variable bigger than $v_k$. This has to stop eventually: either there are no more variables, or some assignment-set is of size 1, or some variable has domain of size 1, or the last considered assignment-set contains assignments to the smallest and the biggest variables (in the order $<$). It is easy to observe that in all cases we have a planar drawing as required in Definition 4.18. If there are some variables left, we continue in the same way. □

Next we show that cross-free VCSPs over 3-element domains with assignments sets of size at most 2 may be neither Knuth-nested nor Knuth-co-nested.

**Example 4.22.** Take four variables $x, y, z, w$ with the domain $\{0, 1, 2\}$, and sets $A_1 = \{\langle x, 0 \rangle, \langle y, 0 \rangle\}, A_2 = \{\langle y, 1 \rangle, \langle z, 0 \rangle\}, A_3 = \{\langle x, 1 \rangle, \langle z, 1 \rangle\}, A_4 = \{\langle y, 2 \rangle, \langle w, 0 \rangle\}$. This instance is cross-free (in fact laminar), but is neither Knuth-nested nor Knuth-co-nested.

## 5. Conclusions

We have studied hybrid reasons for tractability for optimisation problems that can be cast as Valued Constraint Satisfaction Problems (VCSPs), or equivalently Markov Random Fields (MRFs) or Min-Sum problems. These are reasons for tractability that do not follow from the restriction on the functions (such as submodularity) or from the restriction on the structure of the instance (such as bounded treewidth).

Firstly, we have studied binary VCSPs (also known as pairwise MRFs). In the CSP and Max-CSP case, we have obtained a complete dichotomy concerning the tractability of problems defined by placing restrictions on the possible combinations of binary costs in triangles of variable-value assignments. In the case of finite-valued and general-valued VCSP, we have obtained complete dichotomies with respect to equivalence classes which naturally follow from the total order on the valuation structure. We have shown that the joint-winner property and maximum (weighted) matching are the only non-trivial tractable classes.

Secondly, we have studied non-binary VCSPs. We have presented a novel class of optimisation problems that can be solved efficiently using flow techniques. The new class is defined as problems with convex functions over a cross-free family of variable-value assignments. We have shown that neither of the two conditions on its own is sufficient for tractability. Moreover, over Boolean domains, we have managed to extend the new class using the idea of renamability.

We have left open one special case, namely the tractability of cross-free non-convex VCSPs with assignment-sets of size at most 2 and domains of size at most 3. (Assignment-sets of size 3 make the problem intractable even for Boolean domains, and assignment-sets of size 2 over Boolean domains have been shown tractable.)

## Acknowledgments

Martin Cooper is supported by ANR Projects ANR-10-BLAN 0210 and 0214. Stanislav Živný is supported by a Junior Research Fellowship at University College, Oxford.